\documentclass[structabstract]{aa}
\usepackage{graphicx}
\usepackage{txfonts}
\usepackage{dcolumn}
\usepackage{natbib}
\begin{document}

\title{Composition of the galactic center star cluster}
\subtitle{Population analysis from adaptive optics narrow band
  spectral energy distributions}

\author{R. M. Buchholz \inst{1}
       \and
       R. Sch\"odel \inst{2,1}
       \and 
       A. Eckart \inst{1,3} 
       }

\institute{I. Physikalisches Institut, Universit\"at zu K\"oln,
           Z\"ulpicher Str. 77, 50937 K\"oln, Germany\\
	   \email{buchholz,eckart@ph1.uni-koeln.de}
	   \and Instituto de Astrof\'isica de Andaluc\'ia (IAA)-CSIC, Camino Bajo de
	   Hu\'etor 50, E-18008 Granada, Spain\\
	   \email{rainer@iaa.es}
	   \and
           Max-Planck-Institut f\"ur Radioastronomie, 
           Auf dem H\"ugel 69, 53121 Bonn, Germany
           }

\date{Received xx.xx.2009, accepted xx.xx.2009}

\abstract{The GC is the closest galactic nucleus, offering the unique
  possibility to study the population of a dense stellar cluster
 surrounding a SMBH.}{The goals of this work
  are to develop a new method to separate early and late type
  stellar components of a dense stellar cluster based on narrow band filters, to apply
  it to the central parsec of the GC, and to conduct a population
  analysis of this area.}{We use AO assisted observations
obtained at the ESO VLT 
in the NIR H-band and 7 intermediate bands covering the NIR K-band.
A comparison of the resulting SEDs with a blackbody of variable 
extinction then allows us to determine the presence and strength 
of a CO absorption feature to distinguish between early and late type stars.}
{The new method is suitable to classify K giants (and later) as
  well as B2 main sequence (and earlier) stars which are brighter than
  15.5 mag in the K band in the central parsec. 
Compared to previous spectroscopic investigations that are limited to 13-14 mag, 
this represents a major improvement in the depth of the
  observations as well as reducing the needed observation time. Extremely red objects and foreground sources can also be removed 
from the sample reliably. 
Comparison to sources of known classification indicates that the
method has an accuracy of better than $\sim$87\%.
We classify 312 stars as early type candidates out of a sample of 5914 sources. 
Several results such as the shape of the KLF and the spatial distribution 
of both early and late type stars confirm and extend previous works. 
The distribution of the early type stars can be fitted with a steep
  power law ($\beta_{1''} = -1.49 \pm 0.12$, alternatively with a broken
  power law, $\beta_{1-10''} = -1.08 \pm 0.12$, $\beta_{10-20''} = -3.46
  \pm 0.58$, since we find a drop of the early type density at $\sim$10'').
We also detect early type candidates outside of 0.5 pc in significant
  numbers for the first time. The late type density function shows an inversion in the inner 6'',
  with a power law slope of $\beta_{R<6''} = 0.17 \pm 0.09$.
The late type KLF has a power law slope of 0.30$\pm$0.01,
closely resembling the KLF obtained for the bulge of the Milky Way. 
The early type KLF has a much flatter slope of ($0.14 \pm 0.02$).
Our results agree best with an in-situ star formation scenario.}{}

\keywords{Galaxy: center - stars: early-type - stars: late-type - infrared: stars}

\maketitle
\section{Introduction}
 The Galactic center (GC) contains the densest star cluster in the
 Galaxy with a $\sim$$4.0 \times 10^6$ M$_{\sun}$ supermassive black
 hole at its dynamical center
 \citep{eckart2002,schoedel2002,schoedel2003,ghez2003,ghez2008,gillessen2008}. Around
 the black hole, the projected distribution of the stars can best be
 described by a broken power law (break radius R$_{break} = 6''.0 \pm
 1''.0$, all values given here are projected radii), with a power-law
 slope of $\Gamma = 0.19 \pm 0.05$ within the break radius and $\Gamma
 = 0.75 \pm 0.10$ towards the outside of the cluster
 \citep{schoedel2007}. The Galactic center is located at a distance of
 $\sim 8.0$ kpc
 \citep{reid1993,eisenhauer2005,groenewegen2008,ghez2008,gillessen2008}
 from the sun. We will adopt this value throughout this work.

 The stellar composition of the cluster depends on the distance to the
 center. This has been first observed as a drop in CO absorption
 strength towards Sgr A* in seeing limited observations
 \citep{allen1990,sellgren1990,haller1996}. Two explanations have been
 discussed for this: a significantly lower density of late type stars
 in the central few arcseconds and/or the presence of a large number
 of bright early type stars. Adaptive optics assisted observations
 with high spatial resolution have shown that there is indeed an
 increased number of early-type stars in this region,while the
 relative number of late type stars decreases
 \citep{genzel2003,eisenhauer2005,paumard2006,lu2008}. Several authors
 have tried to explain this finding by collisions between stars (or
 between stellar mass black holes and stars), which may lead to the
 destructin of the envelopes of giant stars in the central region
 \citep{davies_benz1991,bailey_davies1999,rasio_shapiro1990,davies1998,alexander1999,dale2008}.\\ Several
 stellar populations have been detected in the central parsec: the
 oldest observable objects that make up the bulk of the visible
 sources outside of the innermost few arcseconds are old, metal-rich
 M, K and G type giants with ages of $1-10 \times 10^9$ years. The
 helium burning {\em red clump} sources are also present, althought
 they have not been discussed in detail until recently
 \citep{maness2007}, because older works on the stellar population did
 not reach the necessary lower magnitude limit.

 A number of intermediate-bright (mag$_K \sim 10-12$) stars that are
 now on the AGB \citep{krabbe1995,blum1996,blum2003} have been
 produced by a star formation event $\geq 100 \times 10^6$ years
 ago. These stars can be distinguished from late type giants by the
 H$_2$O absorption bands in their spectra that are detectable even at
 low spectral resolution. Very few supergiants like IRS 7 are also
 present in the central parsec.  

Several objects with featureless, but very red spectra have also been
detected \citep{becklin1978,krabbe1995,genzel1996}, namely IRS1W, 3,
9, 10W and 21. With high-resolution imaging, most of these sources
have been resolved. They are mostly associated with the mini-spiral,
and can be interpreted as young and bright stars with rapid mass loss
interacting with the interstellar medium in the GC, so called bowshock
sources (\cite{tanner2002,tanner2003,geballe2004}, see also \cite{perger2008} ).

\begin{table}[!b]
\caption{\small Details of the observations used for this work. N is
  the number of exposures that were taken with a given detector
  integration time (DIT). NDIT denotes the number of integrations that
  were averaged online by the read-out electronics during the
  observation. The scale of all observed images is 0.027'' per pixel.}
\label{TabObservations}
\centering       
\begin{tabular}{l l l l l l}
\hline\hline
date & $\lambda_{central}$[$\mu$m] & $\Delta \lambda$ [$\mu$m] & N & NDIT & DIT[sec] \\
\hline
29 April 2006 & 1.66 & 0.33 & 31 & 28 & 2 \\
09 July 2004 & 2.00 & 0.06 & 8 & 4 & 36 \\
12 June 2004 & 2.06 & 0.06 & 96 & 1 & 30 \\
12 June 2004 & 2.24 & 0.06 & 99 & 1 & 30 \\
09 July 2004 & 2.27 & 0.06 & 8 & 4 & 36 \\
09 July 2004 & 2.30 & 0.06 & 8 & 4 & 36 \\
12 June 2004 & 2.33 & 0.06 & 99 & 1 & 30 \\
09 July 2004 & 2.36 & 0.06 & 8 & 4 & 36 \\
\hline
\end{tabular}
\end{table}
In the central $\sim$0.5\,pc there exists yet another distinct stellar
population: massive, young stars created in a starburst $3-7 \times 10^6$
years ago \citep[e.g.,][]{krabbe1995}. These stars can be found, e.g.,
in the IRS\,16 and IRS\,13 associations
\citep[e.g.,]{eckart2004,maillard2004,lu2005}.The brightest of those
young stars have been described as stars
in a transitional phase between O supergiants and Wolf-Rayet stars
(WN9/Ofpe according to e.g. \cite{allen1990}), with high mass-loss
during this phase \citep{najarro1994,krabbe1995,morris1996,najarro1997,paumard2001,moultaka2005}. These stars
account for a large part of the luminosity of the central cluster and
also contribute half of the excitation/ionizing luminosity in
this region \citep{rieke1989,najarro1997,eckart1999,paumard2006,martins2007}. Recently,
  \cite{muzic2008} have identified a co-moving group of highly
  reddened stars north of IRS\,13 that may be even younger
  objects.\\
Besides the most massive early type stars, a large number of OB stars
with masses of $\sim$10-60 M$_{\sun}$ have been examined by
\cite{levinbeloborodov2003,genzel2003,paumard2006,lu2008}. At least
50\% of the early-type stars in the central 0.5\,pc appear to be
located within a clockwise (in projection on the sky) rotating disk,
which was first detected by \cite{levinbeloborodov2003}. Later,
\cite{genzel2003,paumard2006} claimed the existence of a second,
counter-clockwise rotating disk. A very detailed analysis by
\cite{lu2008}, based on the fitting of individual stellar orbits,
shows only one disk and a more randomly distributed off-disk
population (with the number of stars in the disk similar to that on
random orbits). \cite{bartko2008}, on the other hand, claim they at
least observe a counter-clockwise structure that could be a strongly
warped, possibly dissolving second disk. In the immediate vicinity of
Sgr\,A*, there is yet another distinct group of stars, which form a
small cluster of what appear to be early B-type stars
\citep{eckart1999,ghez2003,eisenhauer2005}. These so-called
``S-stars'' stars are on closed orbits around Sgr A*, with velocities
of up to a few thousand km/s and at distances as close as a few
lightdays \citep{schoedel2003,ghez2003,ghez2005,eisenhauer2005,ghez2008,gillessen2008}. Their
orbits have been used to determine the mass of the black hole and the
distance to the GC.\\
How exactly star formation can take place in the central parsec
under the observed conditions is still a debated issue. Classical star
formation from gas of the observed density is severely impeded by the
tidal shear exerted by the black hole and the surrounding dense star
cluster \citep{morris1993}. Two scenarios are being discussed to
explain the presence of the early type stars: 
\cite{genzel2003,goodman2003,levinbeloborodov2003,milosavljevic_loeb2004,nayakshin_cuadra2005,paumard2006}
suggest a model of in-situ star formation, where the infall and
cooling of a large interstellar cloud could lead to the formation of a
gravitationally unstable disk and the stars would be formed directly
out of the fragmenting disk. An alternative scenario has been proposed by
\cite{gerhard2001,mcmillan_portegieszwart2003,portegieszwart2003,kim_morris2003,kim2004,guerkan_rasio2005}
with the {\em infalling cluster scenario}, where the actual star formation takes place
  outside of the hostile environment of the central parsec. Bound,
  massive clusters of young stars can then be transported towards the
  center within a few Myr (dynamical friction in a massive enough
  cluster lets it sink in much more rapidly than individual stars, see
  \cite{gerhard2001}). Recent data seem to favor continuous, in-situ
  star-formation (e.g. \cite{nayakshin_sunyaev2005,paumard2006,lu2008}\\
\begin{figure}[!t]
\centering
\includegraphics[width=\textwidth,angle=-90, scale=0.35]{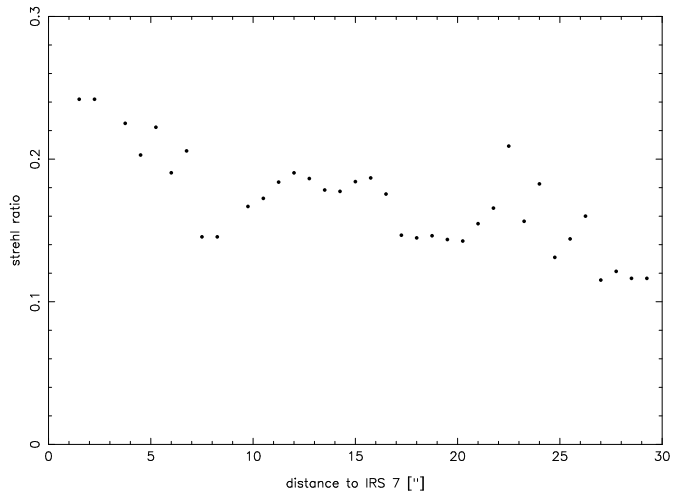}
\caption{\small The strehl ratio, a value that measures the deviation
  of the PSFs of the sources from an ideal PSF, decreases with the
  distance to the guide star.}
\label{Figstrehl}
\end{figure}

The existence of the S stars so close to Sgr A* is yet another matter, known as the ''paradox of youth'' \citep{ghez2003}. Two explanations are discussed for
the presence of these stars, though neither is satisfactory: formation
out of colliding or interacting giants
\citep{eckart1993,genzel2003,ghez2003,ghez2005} or scattering from
the disk of young stars \citep{alexanderlivio2004}.

Detailed spectroscopic studies of the stellar population in the
central parsec have so far only been conducted in the innermost few
arcseconds and on small areas in the outer regions of the cluster (see
\cite{ghez2003,ghez2008,eisenhauer2005,paumard2006,maness2007}). Here
the main limitation is that the high surface density of sources in the
GC forces the observers to use high spatial resolution observations in
order to be able to examine all but the brightest stars. However, the
field-of-view of integral field spectrometers is quite small at the
required angular resolutions (e.g., between $3''\times 3''$ and
$0.8''\times0.8''$ for the ESO SINFONI instrument). The aim of this work
is to provide a reliable method that allows the classification of
several thousands of stars as late or early type in the central parsec
down to a magnitude limit of 15.5. These data will then be used to
obtain constraints on the stellar population and examine the
distribution of early and late type stars. The distribution of very
red objects will also be addressed. For this, information about the
spectra of the stars is necessary. In \S \ref{SectObs}, we present our
method of intermediate band imaging as well as the photometry and
calibration. The disadvantage of this method is the very low spectral
resolution (only 7 datapoints over the K band and one H band point)
that makes it impossible to detect any but the broadest spectral
features. Also, line-of-sight velocities cannot be measured this
way. The big advantage is the possibility to cover a large field of
view of 40''$\times$40'' in a very time efficient way. The data analysis in
  \S \ref{SectPhot}-\ref{SectClassification} will be followed by a population analysis in
\S \ref{SectResults}. We summarize and discuss the implications of our
results in \S \ref{SectDiscussion}.

\begin{table}[!b]
\caption{\small Early type calibration stars, used for the primary
  calibration (names, types and K band magnitudes according to
  \cite{paumard2006}, approximate values for T$_{eff}$ from {\em
    Allen's Astrophysical Quantities})}
\label{TableCalstars}
\centering       
\begin{tabular}{l l l l r}
\hline\hline
name & ID & mag K & type & T$_{eff}$\\
\hline
E69 & 61 & 11.32 & early &  \\
E55 & 195 & 12.45 & B0-1I & $\sim 20000$\\
E47 & 219 & 12.50 & B0-3I & $\sim 18000$\\
IRS 16SSE1 & 148 & 11.90 & O8.5-9.5I & $\sim 32000$\\
IRS 16SSE2 & 199 & 12.10 & B0-0.5I & $\sim 20000$\\
IRS 16SSW & 71 & 11.45 & O8-9.5I & $\sim 32000$\\
IRS 33N & 58 & 11.22 & B0.5-1I & $\sim 20000$\\
E22/W10 & 220 & 12.73 & O8-9.5I/III & $\sim 32000$\\
E25/W14 & 215 & 12.58 & O8.5-9.5I & $\sim 32000$\\
E43 & 126 & 12.10 & O8.5-9.5I & $\sim 32000$\\
E53 & 254 & 12.31 & B0-1I & $\sim 20000$\\
\hline
\end{tabular}
\end{table}
\section{Observation and data reduction}
\subsection{Observation}
\label{SectObs}
The observations used here were carried out with the NAOS-CONICA (NACO)
instrument at the ESO VLT unit telescope 4 on Paranal in June/July
2004 and April 2006 (programs 073.B-0084(A), 073.B-0745(A),
077.B-0014(A)). We used an H band broadband filter and seven
intermediate band filters (see Tab.\ref{TabObservations}).\\
The seeing varied for the different observations, within a range of
0.5 to 1.3''. The bright supergiant IRS 7 located about 6'' north of
Sgr A* was used as the guide star
for the adaptive optics (AO) correction, using the infrared
  wavefront sensor. The
sky background was sampled by taking several dithered exposures of a
dark cloud near the GC, 713'' west and 400'' north of Sgr A*, a region
largely devoid of stars. A rectangular dither pattern was used for
most observations, while some were randomly dithered. All
images were flatfielded, sky subtracted and corrected for dead/hot
pixels.\\
In order to be able to separate early and late type sources using the
method described in \S \ref{SectClassification}, the
photometry has to be accurate enough to clearly identify the feature
used for the classification (see \S \ref{SectFirstfit}). This means
that the typical photometric error should be much lower than the
typical depth of the classification feature (see \S \ref{SectFirstfit}
for an estimate of the required accuracy). When observing
a large field of view like in this case, a good AO correction can only
be achieved within the {\em isoplanatic patch}, a region of $\sim$10-15''
for our dataset. This effect leads to a decrease of the Strehl ratio
towards the outer regions of the field (see Fig.\ref{Figstrehl}). The
values shown in this figure were computed with the {\em strehl}
algorithm in the ESO eclipse software package\footnote{see
  N. Devillard, ''The eclipse software'', The messenger No 87 - March
  1997, publicly available at {\em
    http://www.eso.org/projects/aot/eclipse/distrib/index.html}} from the PSFs determined in 12
$\times$ 12 subimages of the IB227 image (with the same PSF stars that
were used in the photometry). The parameters of the
telescope (like the aperture) were also taken into account. The Strehl
values exhibit a clear trend towards lower
values at larger projected distances from the guide star.   
Sources outside of the isoplanatic patch are elongated towards IRS
7. This is a problem when using PSF fitting photometry, while aperture
photometry (which is less dependent on the shape of the PSF) is faced with
the problem of crowding.

\begin{figure}[!t]
\centering
\includegraphics[width=\textwidth,angle=-90, scale=0.35]{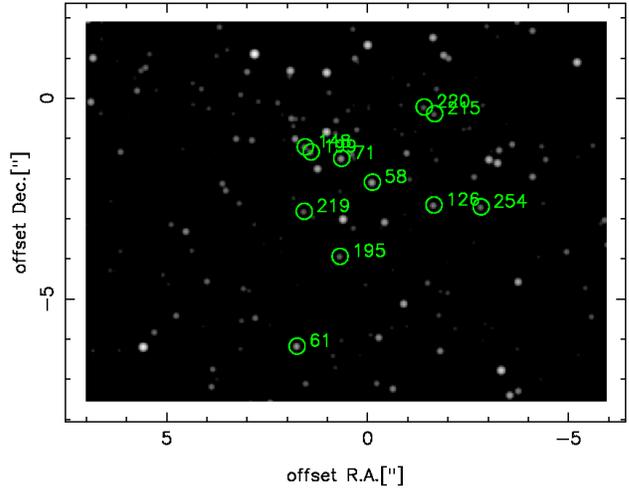}
\caption[Calibration stars]{\small Sources used for the primary calibration. Numbers
  correspond to the numbers in the common list, as also shown in Tab.\ref{TableCalstars}.}
\label{FigCalstars}
\end{figure}

\subsection{Photometry}
\label{SectPhot}
In order to counter the aforementioned problems and to achieve reliable relative
photometry over the entire FOV which is considerably larger than the
isoplanatic patch, a two-step deconvolution process was
used. Deconvolution is a good way to reduce source confusion in a
crowded field, but it can only be applied satisfactorily if the PSF
is very well known and uniform over the whole field. This process is
described in detail in \cite{schoedel2008}, who are using the same photometry.\\
The individual lists of detected stars in the filters were merged to a
common list of stars detected in all 8 filters. The number of common
sources was limited by the image with the lowest quality to
5914. This number includes almost all brighter sources (mag$_{Ks} \leq$
16), but several extended sources (like the bowshock sources in the
northern arm of the minispiral) were not detected in all filters with
sufficiently low photometric and position 
\begin{figure*}[!t]
\centering
\includegraphics[width=\textwidth,angle=-90, scale=0.7]{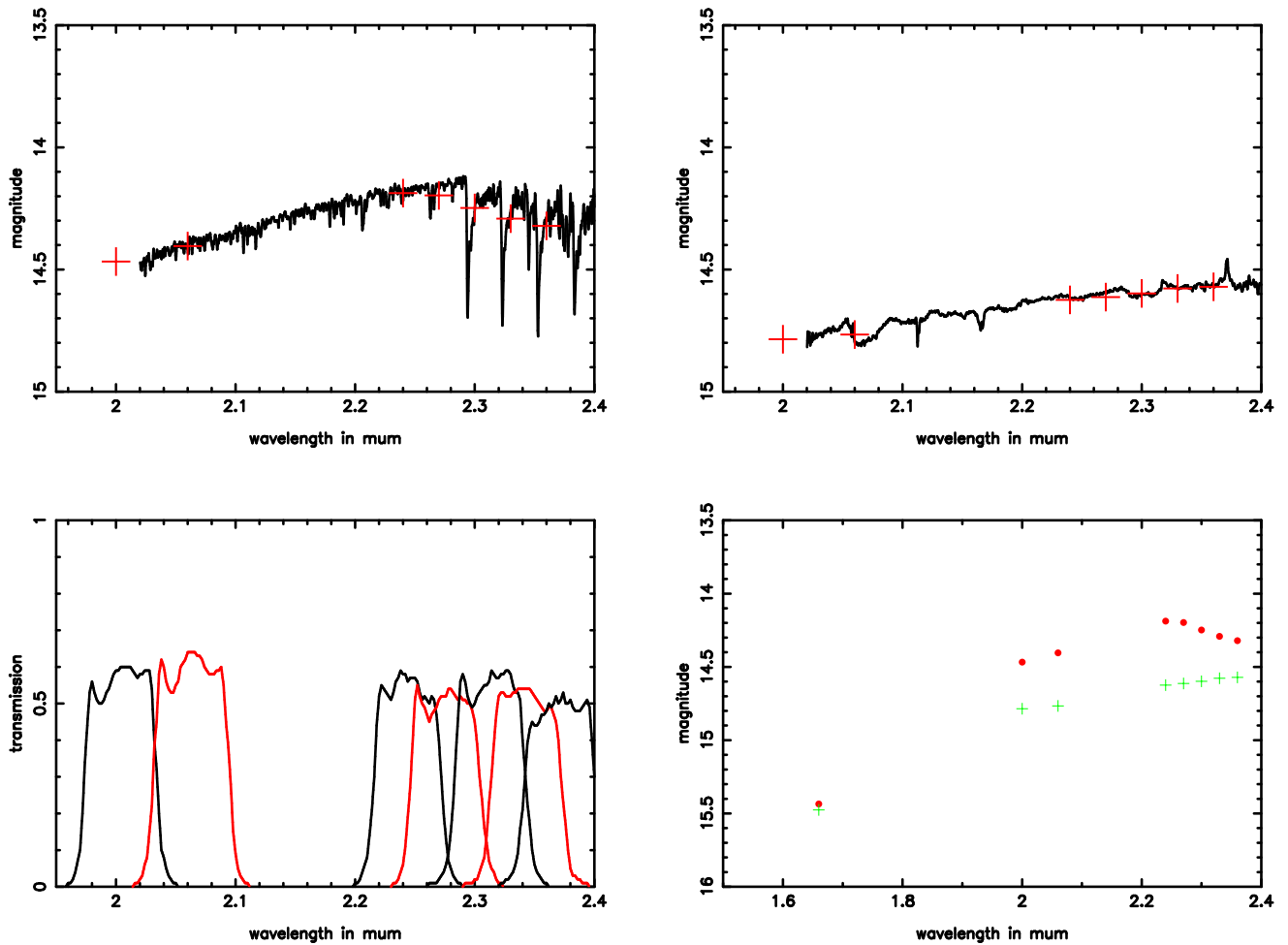}
\caption{\small Conversion of continuous spectra into template SEDs. Upper left:
  K4.5 giant spectrum, the crosses indicate the low resolution SED,
  $A_K = 3.3$ mag. Upper right: B0 main sequence spectrum, $A_K = 3.3$ mag. Lower left:
  transmission curves of IB filters. Lower right: K4.5III and
  B0V template SEDs with added H band datapoint calculated from effective
  temperature and extinction.}
\label{FigWHspectra}
\end{figure*} 
uncertainties.   

\subsection{Primary Calibration}
The primary calibration done here served a dual purpose: on the one
hand, the measured counts for each source in each band were converted
into a magnitude (absolute calibration). On the other hand, the bands
had to be calibrated relative to each other to ensure smooth spectral
energy distributions (SEDs) as they can be expected at this spectral
resolution (see Fig.\ref{FigWHspectra}). Only very broad
spectral features like CO bandheads and H$_2$O absorption bands are
expected to be observable here. Even the feature
around 2.06 $\mu$m that appears in the B0V spectrum only causes a
decrease of 0.014 mag in that filter compared to the K4.5III spectrum
that lacks this feature. This can be neglected compared to other
photometric uncertainties.\\
Atmospheric
features can also influence the shape of the measured SEDs and need to
be eliminated. But the most important parameter that controls the
quality of the data turned out to be the AO performance,
which is why we did not use two of the available intermediate band
datasets (2.12 and 2.18 $\mu$m).\\
These goals were achieved by calibrating the common list of sources with 11 known OB
stars (see Tab.\ref{TableCalstars} and Fig.\ref{FigCalstars}) close to
Sgr A*. We adopted the classifications of \cite{paumard2006}, who
provided a list of 90 early type stars in the central parsec. Stars of
this type have some emission lines in the K band (see Fig.\ref{FigWHspectra}
upper right), e.g. the HeI line at 2.058$\mu$m, but these lines are
narrow enough to be negligible compared to the continuum at the
spectral resolution of our data. Thus, the spectra of these stars can be assumed to be featureless, so they can be
described by a blackbody spectrum with an effective temperature of
$\sim$30000 K. The final calibration has only a minimal dependance on the assumed
effective temperature because the Rayleigh-Jeans law is a very good
approximation for the SED of hot stars in the near-infrared
(using 20000 K instead leads to a difference of
only 0.01 mag in the reference magnitudes). Therefore, the same T$_{eff}$ can be used for all calibration
sources.\\ 
Extinction towards the GC is significant even in the
near-infrared. An extinction of $\sim$3.0 mag
\citep{scoville2003,schoedel2008} has been measured towards the
central parsec in the K band. In order to minimize calibration
uncertainties, we used individual extinction values for each
calibration source, taken from the extinction map provided by
\cite{schoedel2008}. Since the
extinction also depends on the wavelength, we used the
\cite{draine1989} extinction law to calculate an extinction value for
each band from the basic value for the K band, i.e. we assumed:
\begin{equation}
A_{\lambda} \propto \lambda^{-1.75}
\label{EqExtlaw}
\end{equation}
Using an extinction law with a different exponent (like
  $\lambda^{-2.0}$ as proposed by \cite{nishiyama2008})
  would lead to a general offset of about 0.5 mag in extinction, but
  neither the relative distribution of the extinction values nor the
  results of our classification would vary, since the same extinction
  law is used again to fit for the individual extinction of each
  source in the classification process.\\
  \begin{figure*}[!t]
\centering
\includegraphics[width=\textwidth,angle=-90, scale=0.7]{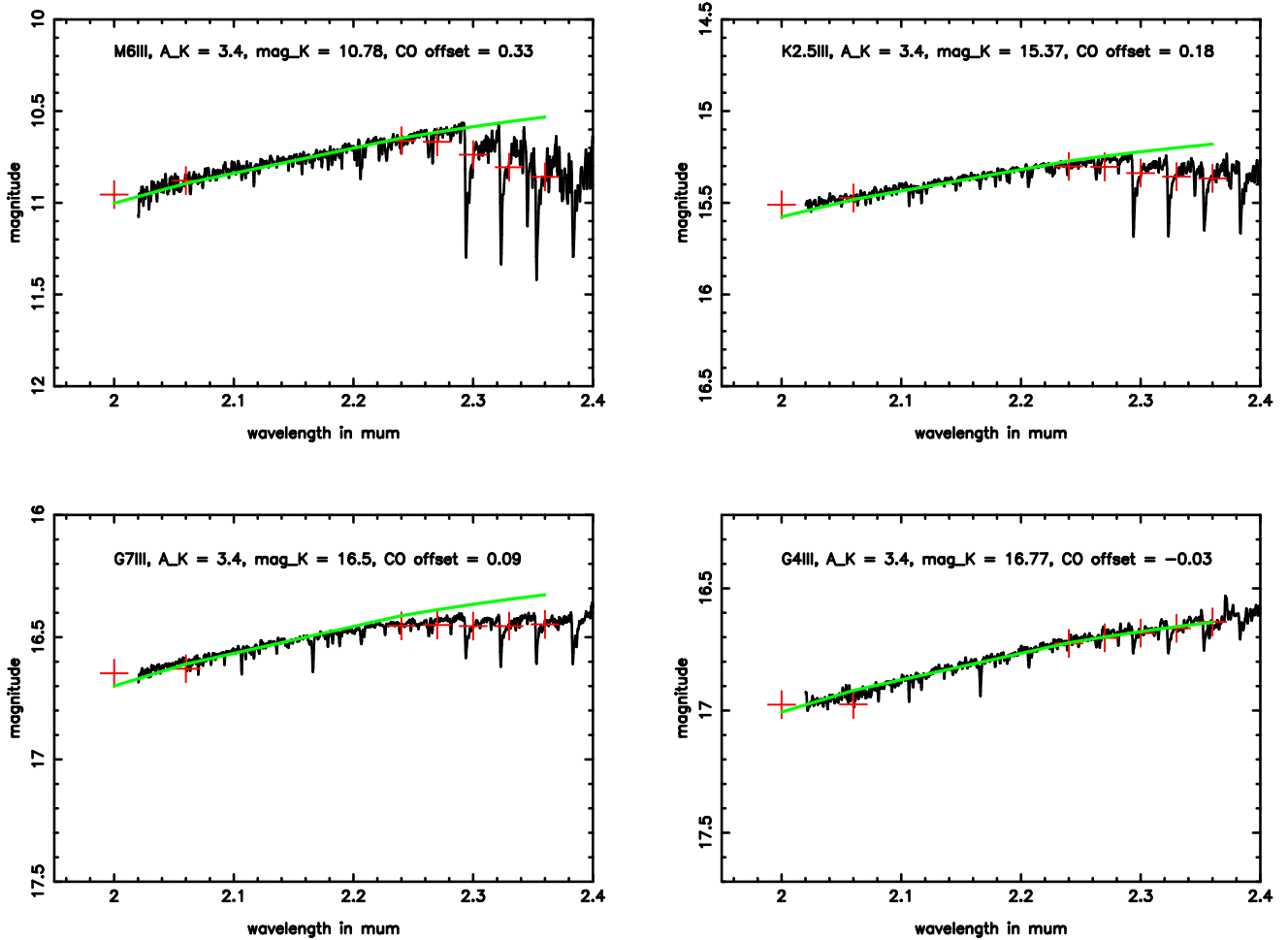}
\caption{\small K band spectra of late type stars expected in the
  GC. The crosses mark the corresponding SEDs and the green solid line
  a fitted extincted blackbody. As can be seen, the CO
  band heads of G giants are not deep and wide enough to have much
  influence on the low-resolution SEDs.}
\label{FigKspectra}
\end{figure*}
The continuous extincted blackbody spectra were converted to 8 point SEDs by
multiplying them with the transmission curve of each filter (see
Fig.\ref{FigWHspectra}, lower left). By comparing these theoretical
SEDs with the measured counts of the calibration stars, a calibration
factor was calculated for each filter to convert the observed counts
of each star into a magnitude. This eliminates the influence of any
atmospheric features, since they should occur in all sources and thus
also in the calibration sources. The magnitude at 2.24$\mu$m was adopted
as the Ks band magnitude of the source, since this band is the closest
available one to the center of the Ks band and not affected by
absorption features. For the classification algorithm and the
analysis, we used extinction corrected magnitudes, i.e. we modified
the measured magnitudes with the difference between the individual
extinction of each source and the average extinction. This step
eliminates the effect that the variable extinction in the observed
region has on the brightness of the sources, which is necessary since
the cutoff used for the classification depends on the brightness of
the source in question. This step also eliminates the effects of
spatially variable reddening on the derived luminosity functions. We assumed
an average extinction in the central parsec of 3.3 mag \citep{schoedel2008}.
\begin{equation}
mag_{ext} = mag_{Ks} - A_{Ks} + A_{avg}
\label{EqExtcorr}
\end{equation}
Here, $A_{Ks}$ denotes the measured individual extinction of each
source in magnitudes (in the Ks band), $A_{avg} = 3.3 $ mag is the
average Ks band extinction in the central parsec, $mag_{Ks}$ represents the
measured Ks band magnitude at 2.24 $\mu$m.\\
After the primary calibration, the sources within a few arcseconds of Sgr A*
show SEDs that agree very well with SEDs calculated from template
spectra of typical stars \citep{wallacehinkle1997}. The optimum of the calibration is centered around the region where the primary
calibration sources are located (close to Sgr A*), and not around the
guide star. With increasing distance from the optimum, the SEDs appear to
show systematic deviations correlated over areas of a few arcseconds
size. There are unfortunately not nearly enough known early type stars
over the whole field to extend our primary calibration to the whole
central parsec, so an additional local calibration had to be
introduced that makes use of another class of stars that are easy to
identify and abundant over the whole field: horizontal branch/red
clump (HB/RC) stars. We assume that almost all late type stars with K
band magnitudes between 14.5 and 15.5 are part of this
population. This leaves the problem of telling apart late and early
type stars. The criterion for this is the same that is later used in
the final classification: the CO band depth (CBD).
\subsection{CO band depth as a classification feature}
\label{SectFirstfit}
Since only very broad spectral features are visible at the low
\begin{table}[!t]
\caption{\small Stellar types expected to be observable in the central parsec, considering
the distance and the extinction towards the GC. Values taken from {\em
    Allen's Astrophysical Quantities}, the types shown here are the ones
    presented in that work that fall into the observational limits for the central parsec.}
\label{TabExpectedStars}
\centering       
\begin{tabular}{l r r l r r}
\hline\hline
early types & mag K & T$_{eff}$ & late types & mag K & T$_{eff}$\\
\hline
B3Ia & 11.80 & $\sim$16000 & M6III & $\sim$3200 & 10.78 \\
O6I & 12.12 & $\sim$36000 & M4III & $\sim$3400 & 12.32 \\
O7V & 13.92 & $\sim$36000 & M2III & 3540 & 12.92 \\
B0V & 14.65 & 30000 & M0III & 3690 & 13.57 \\
B2V & 16.03 & 20900 & K4.5III & $\sim$4100 & 14.29 \\
B3V & 16.33 & $\sim$19000 & K2.5III & $\sim$4300 & 15.37 \\
B7V & 17.51 & $\sim$13000 & G8.5III & $\sim$4750 & 16.42 \\
 & & & G7III & $\sim$4900 & 16.50 \\
 & & & G5III & 5050 & 16.62 \\
 & & & G4III & $\sim$5100& 16.77 \\
 & & & G0III & $\sim$5200& 17.27 \\
\hline
\end{tabular}
\end{table}
spectral resolution of our data, it is not possible to
determine the exact spectral type of every single source. The feature
used to distinguish between early and late types is the region beyond
  2.24 $\mu$m, where late type stars show characteristic CO band head
  absorption. At the spectral resolution available here, no
individual lines or band heads are visible, but a broad
feature like the CO absorption alters the shape of the SED in the
corresponding region significantly (see Fig.\ref{FigKspectra}). The presence of CO band heads causes a
significant dip in the SED for wavelengths greater than 2.24
$\mu$m. In general, the presence of this feature leads to the classification of
the source as a late type star with our method, while its absence makes the star an
early type candidate. We did not consider young stellar objects with
CO band head emission in our analysis. If any such objects exist
within our data, they were treated as early-type stars.\\
This feature also sets the limit for the required photometric accuracy:
  if a late type source with a Ks band magnitude of 15.5 is expected to
  show a CO bandhead feature with a depth of 0.1 mag, this
  corresponds to a difference of $\sim$11\% in flux at 2.36 $\mu$m and
  less at shorter wavelengths. Thus, the photometric error should be
  significantly lower than that value, which is achieved by our method
  for most sources. If the error of a single data-point exceeded 15\%,
  the source was excluded completely.\\     
We determined the presence of this feature in the following way: an
extincted blackbody (T$_{eff}$ = 5000 K, but see \S
\ref{SectClassification} for the effective temperatures used in the
final iteration of the fitting process) was fitted to the first 5
data-points ($\lambda \leq 2.27 \mu$m). The extinction was
  varied in a range of 0 to 8 magnitudes in 0.1 magnitude steps. The best fit was chosen based on the reduced
$\chi^2$ of the fit. A third order polynomial was then fitted to the complete SED, with the first 5
data-points replaced by the fitted extincted blackbody to ensure a
smooth fit. Although it might appear that using only the highest
quality filters that are not influenced by possible additional
spectral features (like the H band and the 2.24 $\mu$m filter) would
produce the best results, this would in fact be less reliable than the
method we used here. Since the GC is a very crowded stellar field,
variable AO performance and related variations of the Strehl ratio can
cause variations of the measured fluxes of stars. This can easily
lead to outliers in the data. Therefore, it is safer to use a larger
number of measurements instead of just two or three filters (see \S
\ref{SectClassification} for the special case of AGB stars with
intrinsic H$_2$O absorption features).\\
 For future applications
  of this method, we would prefer to also use the intermediate band
  filter centered at 2.12 $\mu$m. The data available at this wavelength had a
  low Strehl ratio and thus low photometric accuracy. Using the 2.18
  $\mu$m filter is not advisable, since the data would be influenced
  by the strong Br $\gamma$ emission of the minispiral
  (e.g. \cite{paumard2004}). It would also
  be possible to make use of the available narrow band filters to
  probe regions of special interest in the spectra, e.g. Br $\gamma$
  emission or absorption features in early type stars or the 2.20
  $\mu$m NaI absorption feature in late type stars. But for that, a
  very careful background subtraction would be necessary to eliminate
  the strong influence of the minispiral (see
  e.g. \cite{eisenhauer2005,paumard2006}, the same problem occurs
  with spectroscopy). Most important would be achieving
  improved photometric stability and deeper integration.\\
The shape of both the fitted extincted blackbody and the
third order polynomial depend on depend on the local extinction, but
the difference between the extrapolated values of the two fits at 2.36
$\mu$m is only extremely weekly dependent on extinction. We tested
this by artificially reddening a typical late type SED (modeling
extinction values of 1.6-6.5 mag) and applying the fitting algorithm
to it. The resulting difference of the polynomial and the blackbody
varied by less than 0.15\%, with slightly higher values for lower
extinction. Thus, we can assume that this value is independent of the
local extinction within an acceptable margin of error, which means it
can be used as a good measure for the presence and depth of CO band
heads and thus as a classification criterion. But in order to separate
early and late type stars based on this value, which we will term {\it
  CO band depth (CBD)} in the following, a reliable cutoff is needed.
\subsection{Cutoff determination}
\label{SectCutoff}
\begin{figure}[!t]
\centering
\includegraphics[width=\textwidth,angle=-90, scale=0.3]{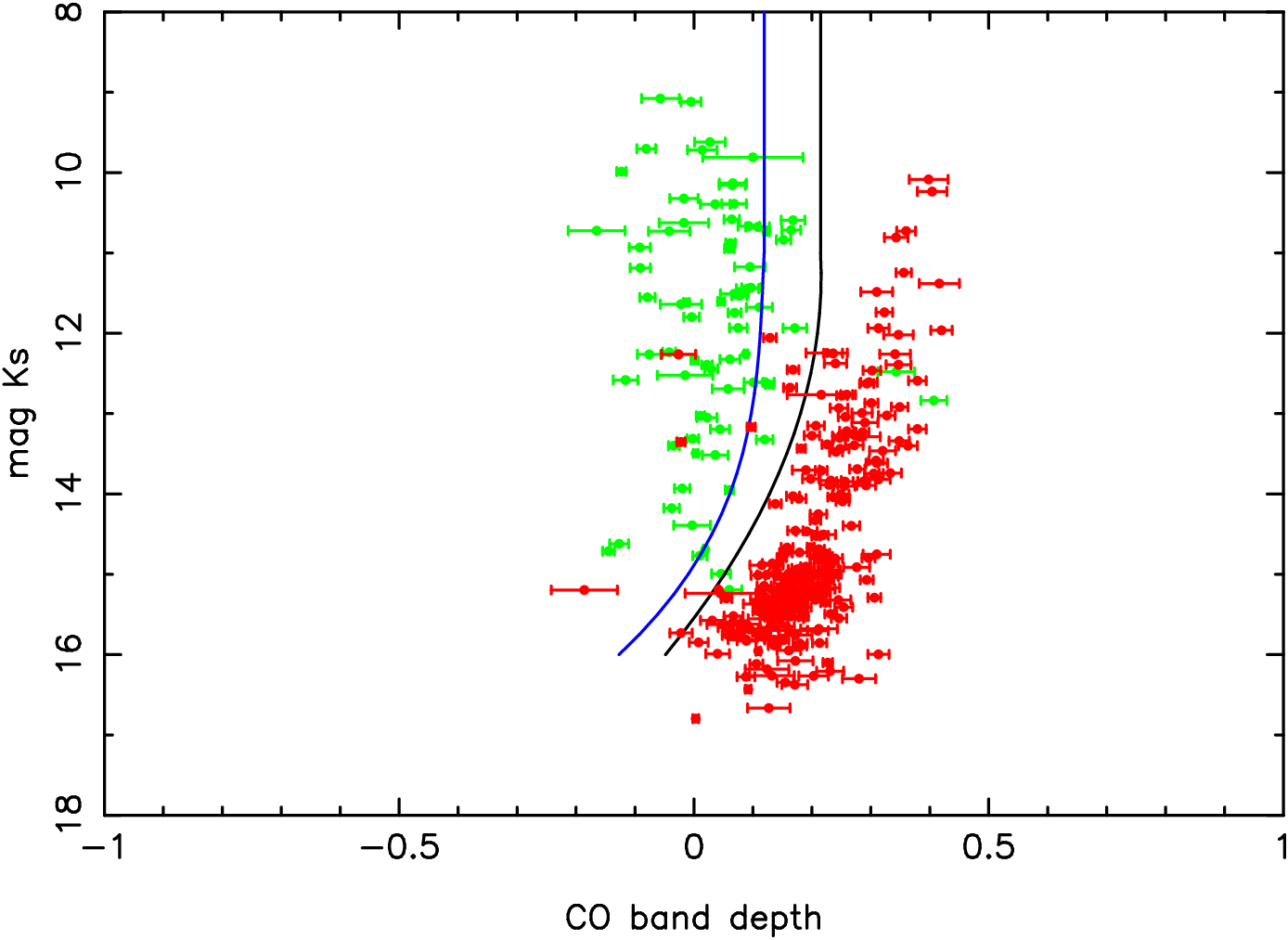}
\caption[CO band depth at 2.36 $\mu$m for reference
  sources]{\small CO band depth (CBD) at 2.36 $\mu$m for known early
  type sources found in this sample (\cite{paumard2006}, green)
  and known late type sources (\cite{maness2007}, red), with the
  cutoff lines plotted in black (derived from early type sources for
  inner 12'') and blue (based on late type sources, used outside of 12''). The cutoffs separate early and late type stars except a few
  outliers. The number of wrongly classified reference sources can be
  used as an estimate for the relative uncertainty of the numbers of
  identified early and late type stars (see \S \ref{SectCompref}).}
\label{FigRefDiffs}
\end{figure}

\begin{figure}[!b]
\centering
\includegraphics[width=\textwidth,angle=-90, scale=0.3]{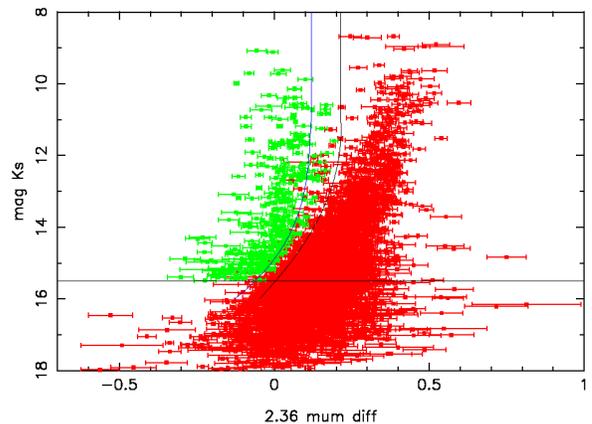}
\caption[CO band depth at 2.36 $\mu$m for sources in the
  central parsec]{\small CO band depth at 2.36 $\mu$m for the
  sources in the central parsec of the GC. Plotted here is the
  difference between the fitted blackbody and the third order
  polynomial at 2.36 $\mu$m. This serves as a criterion for the
  identification of early type (green) and late type (red)
  sources, separated by the inner (black) and outer cutoff (blue solid line).}
\label{FigDiffvsmag}
\end{figure}

\begin{table}[!b]
\caption{\small Stars close to Sgr A* used to create the HB/RC template. All
  these sources were classified as late type by our algorithm, which
  is another self-consistency test.}
\label{TabLocrefstars}
\centering       
\begin{tabular}{l r}
\hline\hline
ID & mag K\\
\hline
3421 & 15.6\\
3369 & 15.1\\
2608 & 15.6\\
2888 & 15.4\\
3896 & 16.1\\
3565 & 15.1\\
4215 & 15.3\\
3253 & 15.7\\
2260 & 15.3\\
2248 & 15.3\\
1440 & 14.6\\
2206 & 14.6\\
4448 & 14.8\\
2158 & 15.0\\
\hline
\end{tabular}
\end{table}
Although one could expect from template SEDs as shown in Fig.\ref{FigWHspectra}
a CBD close to zero for early type sources and
greater than zero for late type sources, which could be separated by a
simple general cutoff, the limited signal-to-noise ratio of the data,
combined with additional sources of uncertainty, such as photometric
errors due to source crowding (see e.g. \cite{ghez2008}) led to
considerable scatter in the CBD feature. Therefore, the CBD value of
early type stars scatter to positive and negative values. But still,
this value can be used as a useful tool to reliably separate early- and
late-type stars statistically.\\
In order to compensate for
these difficulties, we compared the CBD values of the sources classified by \cite{paumard2006,maness2007} that are
also present in our dataset. As can be seen in Fig.\ref{FigRefDiffs},
the early- and late-type sources mostly fall into separate regions
when we plot the stellar magnitude vs.\ the CBD value. However, there
are some (less than 5\%) late-type sources present in the early-type
region and vice versa. This makes it impossible to use a single smooth
cutoff line, let alone a single cutoff value for the whole magnitude
range.\\
Almost all spectroscopically identified
early-type sources are concentrated in the innermost
12.9''. Therefore, two different smooth cutoff lines were calculated:
one that reliably separates all known early and late type sources in
the inner 0.5 pc, and one for the outer regions that encompassed the
known late type sources there (see Fig.\ref{FigRefDiffs}). This
ensures a more reliable detection of early type stars in the outer
region (where the early type density is at best very low. The cutoff
line based on the early-type stars will underestimate the number of
early type stars, just as using the less strict cutoff based on the
late-type stars might overestimate the number of early type stars. Use
of both cutoff lines allows us to estimate the uncertainty of this
method.\\ We were able to achieve a clear separation of the reference
sources with this two-cutoff-method, except a few outliers and noisy
sources (see \S \ref{SectCompref}).\\
The theoretical lower magnitude limit of this method is determined by
the presence of deep enough CO band heads so that the CBD significantly
exceeds the photometric uncertainties. This was determined from the
\cite{wallacehinkle1997} spectra to be the case for M, K and brighter
G giants, corresponding to $\sim$15.5-16 mag at the distance and
extinction towards the GC. Fainter G giants and early type stars
fainter than B2 have almost identical CBD values (see
Fig.\ref{FigKspectra}). The comparison with published spectroscopic
identifications and the shape of the CBD vs. magnitude plot
(Fig.\ref{FigDiffvsmag}) justified the adoption of a magnitude limit
of 15.5.
\subsection{Local calibration}
\label{SectLocCal}
\begin{figure*}[!t]
\centering
\includegraphics[width=\textwidth,angle=-90, scale=0.7]{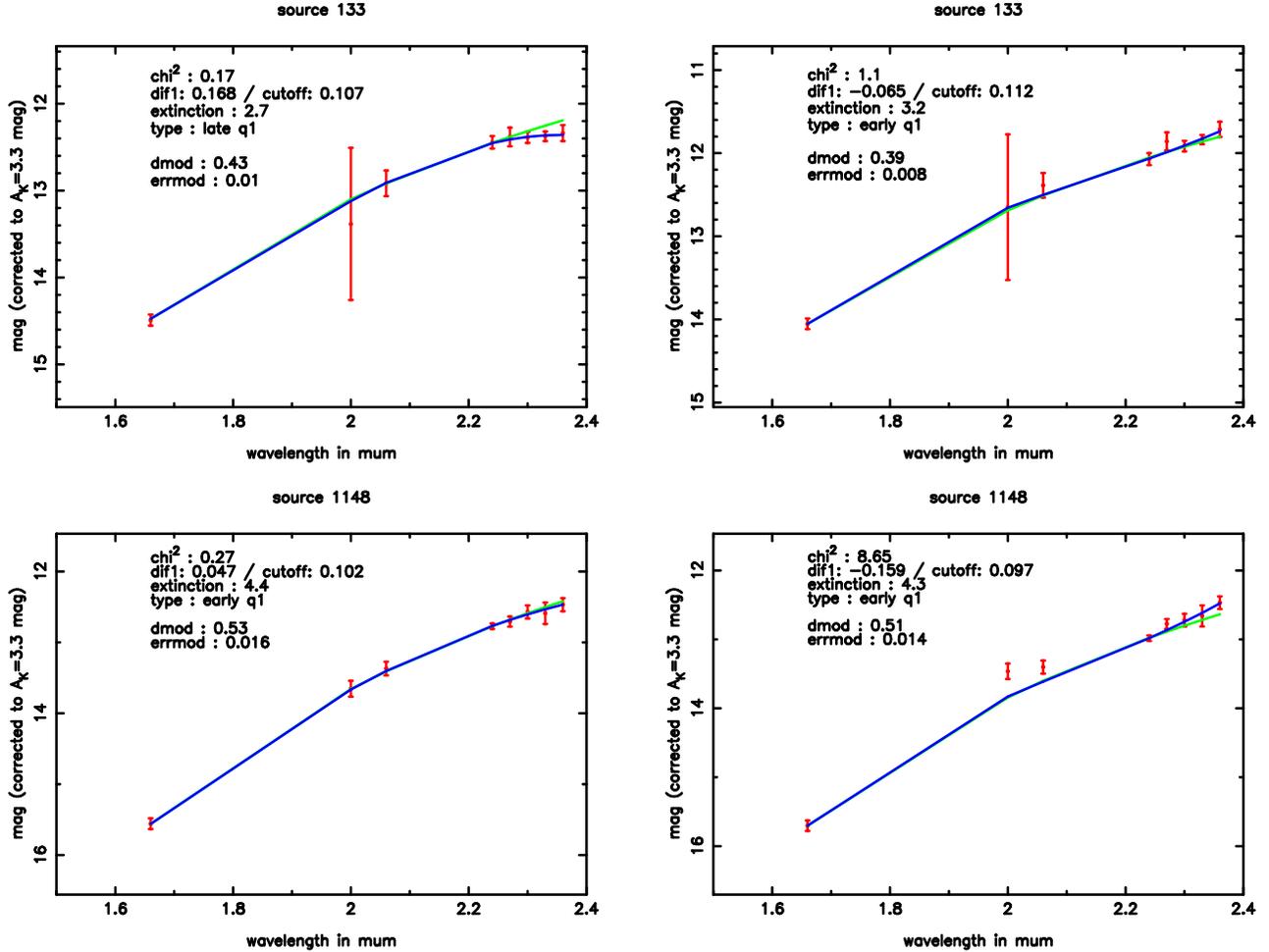}
\caption[Effects of the local calibration]{\small Two examples for the
  effects of the local
  calibration. Upper right: source fitted as early type, but the SED
  is noisy which is reflected in the $\chi^2$. Upper left: same source
  after local calibration, now fitted as late type, with a much better
  fit. Lower right: source fitted as early type, noisy SED, several
  data-points do not agree with the fit, large $\chi^2$. Lower left:
  same source, still fitted as early type, but with much better
  fit.}
\label{Figloccal}
\end{figure*}

With a classification criterion and a reliable cutoff in place, we
were able to apply a local calibration that significantly improved the
results.\\ 
Previously published K band luminosity functions of the central parsec
(e.g. \cite{genzel2003,maness2007,schoedel2007}) have shown that the magnitude
range of 14.5 to 15.5 is dominated by HB/RC stars outside of the
innermost arcsecond. In addition, the density of early type stars
decreases steeply outside of the innermost arcsecond
\citep{genzel2003,paumard2006,lu2008}. So even without knowing the
exact type of every source in this magnitude range, they can be used
as calibration sources assuming a typical late type SED. This template
was determined as the average SED of 14 manually selected late
type reference stars (see Tab.\ref{TabLocrefstars}) close to the optimum of the calibration. These
stars were selected based on the similarity of their SEDs to the
expected shape.\\
After this, the HB/RC sources in the whole field except the innermost
2'' were selected based on two criteria: an extinction-corrected
magnitude between 14.5 and 15.5 and a CBD value above the cutoff. The SEDs of these sources were compared to 
the template HB/RC SED corrected for the determined individual
extinction value. This yielded a calibration factor for each source in
each band. Due to the uniform distribution of HB/RC sources over the
field, these values map the local deviations and can be used to
calibrate the SEDs of all sources. The calibration factors were
checked for outliers first. If a calibration factor exceeded the
median over the closest 20 sources by more than 3 $\sigma$, it was
replaced by that median. On average, 20 sources are contained within
an area of $\sim$2 $\times$ 2'', depending on the position in the
field. This already leads to a spatial resolution of $\sim$2''. The
factors were then processed into 8 calibration maps by smoothing the
raw maps with a Gaussian with a 2'' FWHM (see Fig.\ref{FigCalmaps}).\\
The final cutoff to be used in the actual classification procedure was
calculated in the same way as described in \S \ref{SectCutoff}, but
with the locally calibrated values for the reference sources. This
turned out to be a minor adjustment, so the local calibration process
did not have to be reiterated.\\
This local calibration eliminates most systematic local deviations and
allows a more reliable classification of the sources towards the outer
edges of the field (see \ref{Figloccal}).
\subsection{Source classification}
\label{SectClassification}
The fitting process described in \S \ref{SectFirstfit} was repeated on
the calibrated data. The extinction, CBD, the $\chi^2$ of the fitted
polynomial and the uncertainty of the CBD were calculated for each source in
several iterations (see below). The reduced $\chi^2$ value for the
last five data-points of the polynomial fit was used as a criterion for
the quality of the fit. Using a reduced $\chi^2$ for all eight
data-points makes less sense than the method used here, since, on the
one hand, the first five data-points are replaced by the fitted
blackbody for the polynomial fitting, and on the other hand, the
first data-points do not have a large influence on the classification
anyway. In order to allow a realistic comparison to the cutoff for the final
classification, the uncertainty of the CBD was calculated as the average
root-mean-square deviation of the last five data-points to the
polynomial fit. This value deliberately ignores the individual
photometric uncertainties of the data-points, because sources with noisy SEDs
with large uncertainties for the last data-points that are crucial for the
classification would otherwise have a similar CBD uncertainty as
sources with smooth SEDs with the last data-points close to the fitted
blackbody. This could lead to erroneous classifications and is not desirable.\\
We separated our sample of 5914 stars into the following classes:
\begin{enumerate}
\item Noisy sources: if the reduced $\chi^2$ of a source was higher
  than 1.5, we rated this source as too noisy for classification (334
  sources). The cutoff of 1.5 was chosen because it excludes the 5\% most noisy
  sources.
\item Foreground sources: these stars are not part of the population
  of the central parsec and have to be removed prior to any
  analysis. They are easily recognizable by their low fitted
  extinction value. Every source with a fitted extinction of less than 2
  magnitudes was rated as a foreground source here and excluded. 58
  sources were classified as foreground sources.
\item Very red objects: several very strongly reddened objects (like
  e.g. IRS 3) can be observed in the field. Their SEDs are influenced
  by other effects in addition to their intrinsic stellar features,
  e.g. dust shells and bow-shocks. These sources were also excluded
  from the further analysis since they cannot be compared easily to
  normal early or late type stars. We rated every source with a fitted
  extinction of more than 5 magnitudes as such an object (24
  sources). The high fitted extinction here is just a selection
  criterion that results from our algorithm. The real extinction toward
  the sources cannot been determined because of the unknown intrinsic
  SED of the very red objects.
\begin{figure*}[!t]
\centering
\includegraphics[width=\textwidth,angle=-90, scale=0.35]{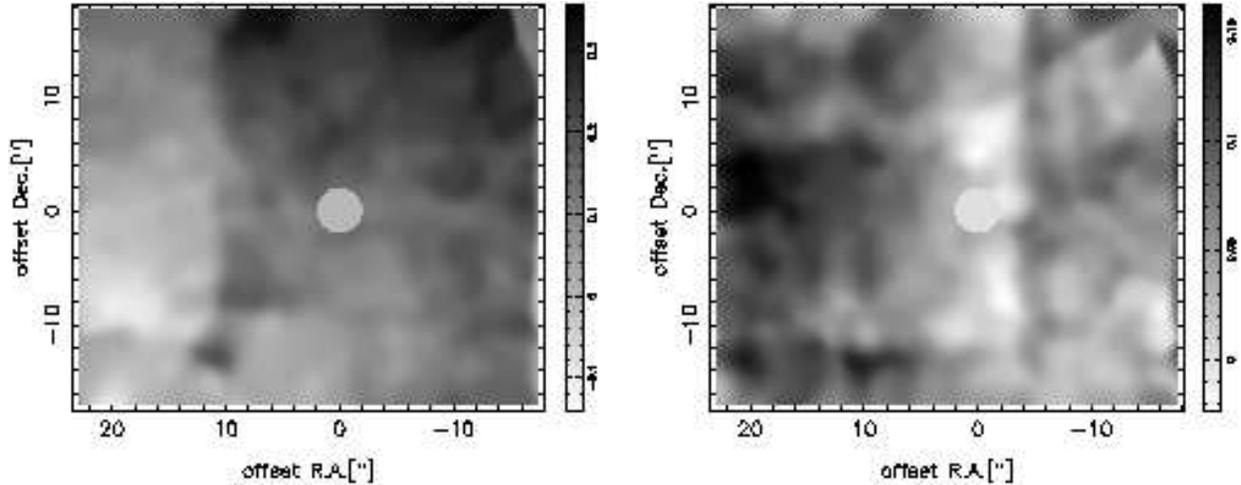}
\caption{\small Examples of the calibration maps used for local
  calibration (see \S \ref{SectLocCal}) in the 8
  bands. Left: 2.00 $\mu$m. Right: 2.33 $\mu$m. A central region with a radius of 2'' was excluded, the
  calibration factors were set to 0 there. These maps were generated
  by comparing a median spectrum of known HB/RC sources close to the
  optimum of the primary calibration with presumed HB/RC sources all
  over the field (see \S \ref{SectLocCal}). Residuals of the
    rectangular dither pattern that was applied in the
    observations can be seen in the calibration maps.}
\label{FigCalmaps}
\end{figure*}
\item Early type sources (quality 1-3): the CBD of every source
  brighter than 16 magnitudes (extinction corrected) was compared to
  the applicable cutoff: sources closer than 11.9'' were compared to
  the inner cutoff, while any source further away than 11.9'' was
  compared to the outer cutoff. Sources with a CBD below the cutoff
  were rated as preliminary early type candidates and fitted again with  T$_{eff}$ =
  30000 K which comes much closer to the actual effective temperatures
  of early type stars than the 5000 K that were
  used in the initial fit. A 30000 K blackbody has a flatter slope in
  this wavelength regime than a 5000 K blackbody, so the fitted
  extinction and the CBD that results from the fit also differ.
  The updated CBD value was compared to the cutoff again. Only sources
  brighter than 15.5 magnitudes (extinction corrected) were
  considered. The extinction fitted in this iteration exceeded the
  previous value by up to $\sim$0.5 due to the flatter slope of the
  blackbody. This explains the higher magnitude cutoff used in the
  first step. A classification of fainter sources is not possible with this
  method, because the depth of the CO band head feature becomes too
  shallow \citep{wallacehinkle1997}. If the CBD was more than 3
  $\sigma$ lower than the cutoff and the SED was not too noisy
  (reduced $\chi^2 <$ 1.5), the source was rated as an
  early type with quality 1 (highest quality, 277 sources). If CBD was
  between 2 and 3 $\sigma$ lower than the cutoff, the source was rated
  as an early type with quality 2 (25 sources). Sources with a CBD 1-2
  $\sigma$ below the cutoff were rated as early type quality 3 (10
  sources).
\begin{figure}[!b]
\centering
\includegraphics[width=\textwidth,angle=-90, scale=0.4]{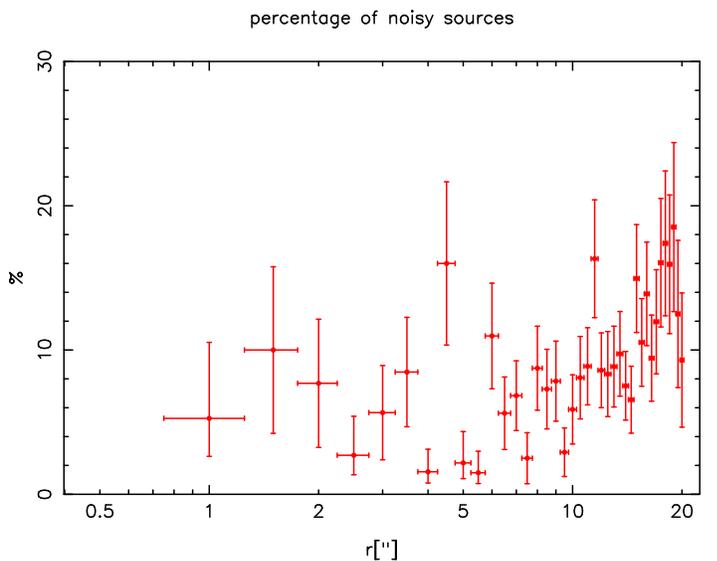}
\caption{\small Azimuthally averaged density of stars with noisy
  SEDs. The density profile is practically flat over the inner 15'',
  with a slight increase further out. This indicates a good photometry and a reliable mechanism to
  identify noisy sources, since no region seems to have an over-density
  of noisy sources. This is also a
  good indication that the local calibration improved the results.}
\label{FigPDensityN}
\end{figure}
\item Late type sources (quality 1-2): all other sources that did not
  meet the cutoff criterion to qualify as an early type candidate were
  fitted again with T$_{eff}$ = 4000 K. This value is typical for the
  expected late type giants. This again resulted in different
  extinction and CBD values. The uncertainties of the CBD and the reduced
  $\chi^2$ of the fit were calculated in the same way as for the early
  type stars. If the CBD exceeded the cutoff by more than 1 $\sigma$
  and the magnitude of the source was brighter than 15.5, the source was
  rated as a late type star, quality 1 (2955 sources), the others were
  rated as late type, quality 2 (2231 sources). The late type quality
  2 sources are neglected in the analysis of the late type population,
  since our method does not provide a clear identification any more.
\item AGB stars: AGB stars are among the brightest sources in the
  field. Due to their prominent H$_2$O absorption feature, they are
  usually fitted with a too flat blackbody/polynomial: the
  absorption feature leads to a higher magnitude at 2.00 and 2.06
  $\mu$m that drags the fitted blackbody and the polynomial
  ''down''. This can lead to a CBD below the cutoff and a classification as an early type star. To counter
  this effect, all sources were checked whether or not these two
  data-points were more than 1 $\sigma$ below the fitted blackbody. If
  that was the case, the fitting was repeated without them. When the
  resulting classification changed, we adopted the updated class as
  the better result. We did not deem it necessary to put these objects
  in a separate sub-category, but we see this modification as a way to
  further improve the classification of early type stars and remove
\begin{figure*}[!t]
\centering
\includegraphics[width=\textwidth,angle=-90, scale=0.7]{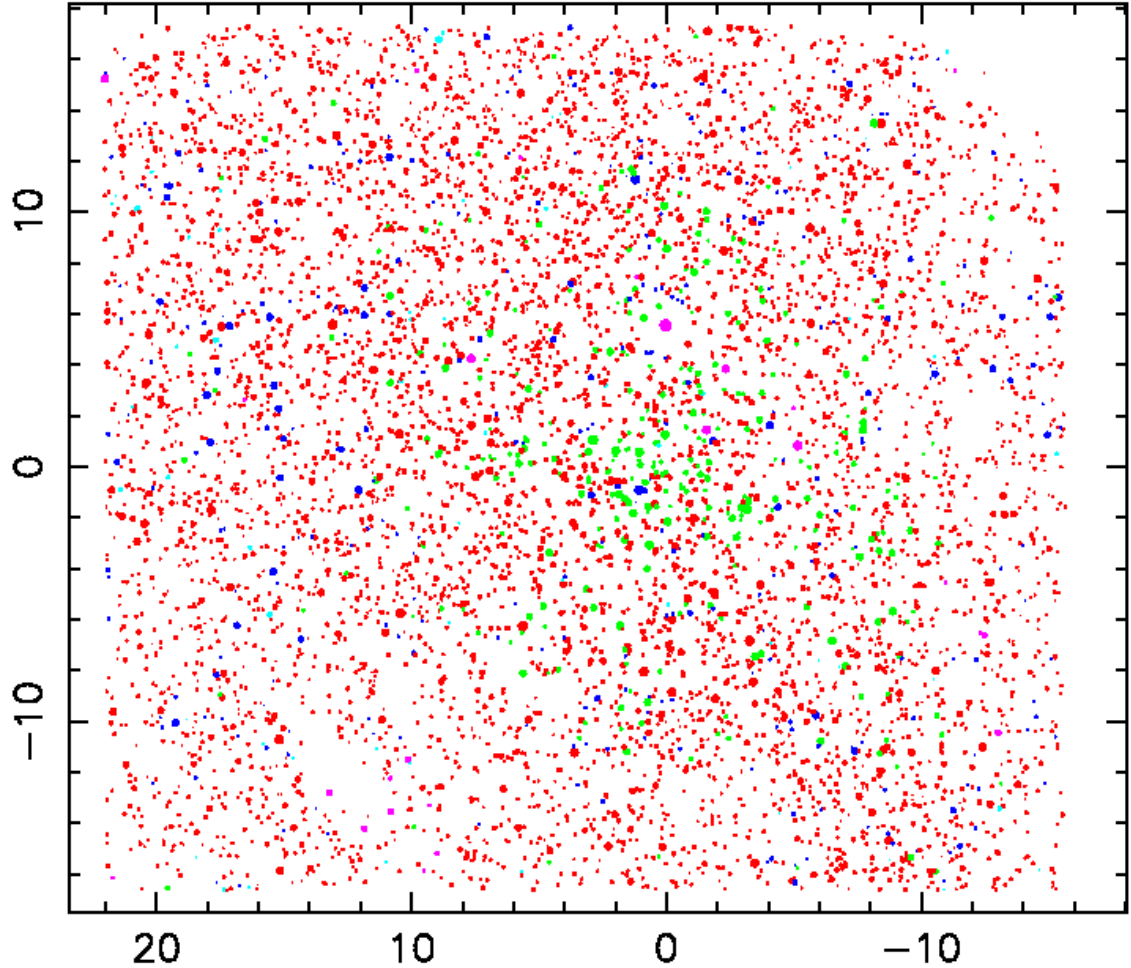}
\caption{\small Map of the stars in the GC. Red circles represent late
  type stars, green circles stand for early type candidates. The blue
  circles are sources not classified, light blue indicates foreground
  sources, while magenta stands for extremely red objects. The radii
  of the circles are linearly dependent on the extinction corrected
  magnitudes of the sources.}
\label{MapEarlyLate}
\end{figure*}
  false candidates. Source 1049 is a good example for this effect,
  especially since it has been identified spectroscopically as a late
  type source by \cite{maness2007}. This addition to the algorithm
  resulted in 8 less candidates for early type stars in total, which
  does not have a big impact on our results as a whole, although it
  has a noticeable effect on the density distribution in the outer region
  of the observed area.
\end{enumerate}
Using only two different values for T$_{eff}$ to fit early and late
  type candidates turned out to be sufficient, since the variations in
  the fitted extinction stayed within $\pm 0.1$mag when we used the
  minimum resp. maximum values for T$_{eff}$ as shown in
  Tab.\ref{TabExpectedStars} for the respective classes of objects. We
  only determined the extinction in steps of 0.1 mag in the first
  place, so we considered this slight uncertainty acceptable.
\section{Results and discussion}
\label{SectResults}
\subsection{Stellar classification}
\label{SectClass}
In total, 3349 of 5914 sources have been classified as either early
type, late type, foreground or very red sources. 2231 of the remaining
sources are too faint to allow a reliable classification with our
method, although they can be assumed to be mostly late type
sources. The SEDs of 334 sources were too noisy and have been excluded
by the classification algorithm.\\ Unless otherwise indicated, all
results shown here are based on the sources with an extinction
corrected magnitude brighter than 15.5 (see
Eq.\ref{EqExtcorr}). Foreground sources have also been excluded (58 in
total), since they do not belong to the population of the central
parsec. Very red objects (24 sources) have also not been included in
the number of early or late type stars, but have been treated as
an extra class of objects. Objects in this class can be of different type:
highly extincted background objects (and thus not very relevant for an
analysis of the central parsec), sources with dusty envelopes,
e.g. IRS\,3 \citep{pott2008}, young stars, whose
strong winds interact with the interstellar medium in the form of
bow-shocks \citep{tanner2002,tanner2003}, or even candidates for young
stellar objects \citep{muzic2008} (examples for the SEDs of these
sources are shown in Fig.\ref{FigEROs}). Tab.\ref{TabIdentifiedStars} lists the
stars in each (sub)category, while Fig.\ref{MapEarlyLate} shows the
spatial distribution of the classified sources.
\subsection{Comparison with spectroscopic results and uncertainty estimation}
\label{SectCompref}
We compared the results of our classification to the lists of
spectroscopically classified sources provided by \cite{paumard2006}
and \cite{maness2007}. The list published by \cite{paumard2006}
contains 90 early type stars, 78 of which were contained in our list
of common sources. This discrepancy can be explained by the nature of
the observations used here (different observation dates for the
individual filters, 2 years between H band and IB observations,
different quality of the datasets), some sources (especially fast
moving objects like S2) were not detected at the same position in
every image, not detected through all filters or excluded due to too
large photometric uncertainties and thus did not make it on the common
list. The same effects are probably relevant for the 329 sources
classified by \cite{maness2007}, while in addition, parts of this
sample are outside of the area covered by our data. We only used the
266 sources that are also present in our dataset for comparison.\\ 67
of the 78 known early type sources were also classified as early type
here, 7 had too noisy SEDs and 4 were classified as late type. Of
these 4 sources, 3 are borderline cases where a clear identification
is very difficult with our method (sources 224, 612 and 1772
resp. E36, E89 and E7 in the Paumard list).
\begin{table}[!t]
\caption{\small Stars classified in the GC using the method described in Sect.\ref{SectClassification}.}
\label{TabIdentifiedStars}
\centering       
\begin{tabular}{l r}
\hline\hline
class & number\\
\hline
early type quality 1 & 277 \\
early type quality 2 & 25 \\
early type quality 3 & 10 \\
late type quality 1 & 2955 \\
late type quality 2 &  2231\\
foreground & 58 \\
very red stars & 24 \\
noisy sources & 334 \\
\hline
\end{tabular}
\end{table}
Source 3778 (E37 in the
Paumard list) shows a clear CO absorption feature. \cite{paumard2006}
classify this source as a potential O8-9 supergiant, but at the same
time give it a K band magnitude of 14.8 and an absolute magnitude of
-3.3. This is inconsistent with our expectations, since such a source
should be at least two magnitudes brighter (see
Tab.\ref{TabExpectedStars}). We therefore ignore this source for the
uncertainty estimation. This leads us to 3 out of 77 sources
classified erroneously and 7 out of 77 sources not classified, which
corresponds to $3.9\%$ respectively $9.1\%$. A few well known sources
like IRS 16SW and IRS 15NE have noisy SEDs, which in the case of IRS 16SW
is probably due to the intrinsic variability of that source. But in general,
noisy SEDs mostly stem from problems with the photometry: here, too faint or
saturated sources are the biggest issues.\\ 
Since the known early type sources are
concentrated in the inner 0.5 pc, these values can be adopted as the
uncertainties of the number of early type stars in the innermost few
arcseconds identified in this work.\\ 258 of the 266 known late type
sources have been classified as late types by our method. The SEDs of
7 sources were too noisy and one was classified as early type (source
363, 96 in Maness list). Source 363 does not show a clear CO feature
despite being bright enough, and a comparison of the Maness and
Paumard lists shows that there is an early type source 0.17'' from its position
(assuming that the positions given in these works use the same reference frame). In the imaging data that we
  have used for this work, there is only a single source present at
  the location of source E87 (Paumard) resp. 96 (Maness). It appears 
  sufficiently isolated to rule out confusion with another
  source. This leads us to the assumption that \cite{paumard2006} and
  \cite{maness2007} are looking at the same source there, but classify
  it differently.\\
 In order to derive an upper limit for the
  uncertainty and thus the confidence in our method, we assume one
  erroneous classification in the area covered by the SINFONI
  observations. In this region (north of Sgr A*, mostly outside of 0.5
  pc), we find a total number of 11 sources classified as early type,
  including the one star of disputed type. If we assume one of these
  classifications to be erroneous, this leads us to an uncertainty of
  $\sim$9\% for our number of early type stars outside of 0.5 pc. We
  consider this an acceptable level of confidence, considering the low
  density of early type stars we measure this far out.\\ To be on the
  conservative side, we adopt this value,$\pm 9\%$, as the uncertainty
  of the total number of all early type stars, i.e. in the entire
  field of view. It has to be considered, however, that the low total
  number of sources available for the determination of this uncertainty
  level limits the confidence in it.
\subsection{Structure of the cluster}
\label{SectStructure}
\begin{figure}[!b]
\centering
\includegraphics[width=\textwidth,angle=-90, scale=0.35]{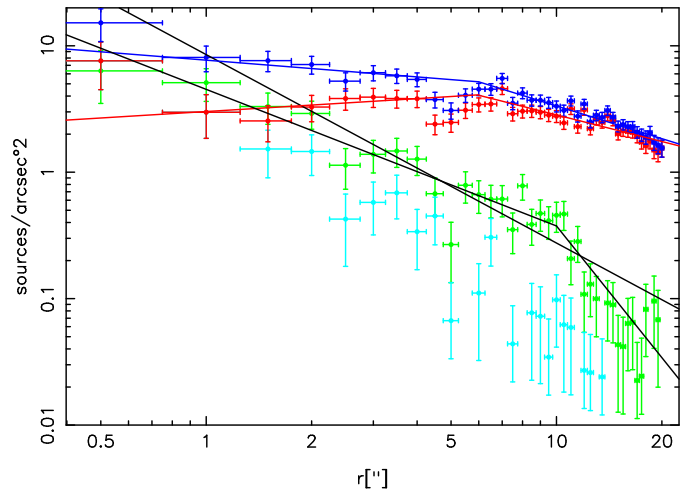}
\caption{\small Azimuthally averaged stellar surface density plotted
  against the distance to Sgr A* for Ks magnitudes mag$_{Ks} <
  15.5$. No completeness correction has been applied here, but the
  data can be assumed to be complete down to mag$_{Ks} \sim $ 15.5
  \citep{schoedel2007}. The green points describe the distribution of
  early type stars, while red stands for late type quality 1 stars and
  dark blue for all detected stars. This also includes stars rated as noisy and
  bright enough late type quality 2 sources. The early type stars given in
  \cite{paumard2006} are shown in light blue for comparison. The solid
  lines indicate the power laws fitted to the data.}
\label{FigPDensity}
\end{figure}

Fig.\ref{FigPDensity} shows the projected stellar density for the
total population, the early and late type stars. For comparison, we
also show the projected density of the early type
stars provided by \cite{paumard2006}. Only stars brighter than 15.5
mag have been considered here, in order to allow a clear separation
of early and late type stars and to make a completeness correction
unnecessary (see \cite{schoedel2007}).\\
The projected density profile of the late type stars is practically
flat within a radius of $\sim$10''. Within the innermost 5'', it can
even be fitted with a power law with a positive slope, i.e. the
projected density increases with the distance to the center. This
flattening or even inversion of the projected surface density profile
of the late-type stars combined with the steeply increasing density of
early-type stars towards Sgr\,A* explains the dip in CO band head
absorption strength found in early spectroscopic observations of low
spatial resolution \citep{allen1990,sellgren1990,haller1996}. We
discuss this in detail in \S \ref{SectGiants}.\\
A dip in the density can be observed at a radius of $\sim$5''
  that has already been observed by \cite{schoedel2007} in deep
  ($mag_{K} \leq 17.5$) star counts. \cite{zhu2008} also find a dip
at 0.2 pc, which corresponds to the 5'' given here and in
\cite{schoedel2007}. This dip is a significant feature in the
  density profiles. It is both present in the late- and early-type
  population, although with a low significance in the latter, due to
  the small number of early-type stars. The cause for this feature
is probably extinction, since there is a ring-like area of high
extinction visible in our extinction map at this distance to the
center (see Fig.\ref{FigExtmap}). Since this dip feature seems to
  appear in two stellar populations that are so different in their age
  and their dynamical state and since it would be very difficult to
  reproduce such a feature in a three-dimensional distribution,
extinction seems to be the most likely explanation.\\
A steep increase of the projected density of early type stars can be observed
towards the center (except in the innermost arcsecond). This also
confirms previous observations. \cite{paumard2006} proposed that the
projected density of the early type stars follows a
$R^{-2}$ power law outside of a sharp inner edge at 1'' (within the
  disk(s), so this value is not directly comparable to our findings), with $R$ as the
projected distance to Sgr A*. \cite{lu2008} also confirmed the
  $R^{-2}$ power law within the clockwise disk. To allow a comparison
  with our values, we fitted the projected density of the known early
  type sources that are contained in our sample with a power las as
  well. This yielded a value of $\beta_{ref} = -1.80 \pm 0.17$. In the
  following, we adopt this value as a reference.\\
\begin{figure}[!t]
\centering
\includegraphics[width=\textwidth,angle=-90, scale=0.4]{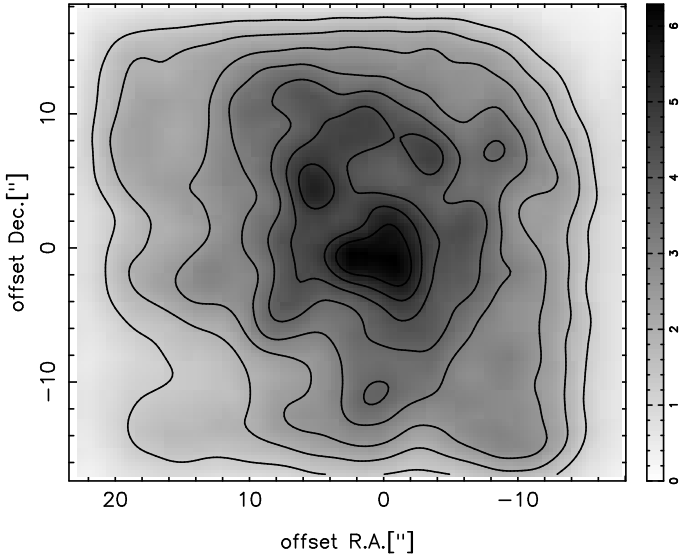}
\caption{\small Stellar surface density of all stars brighter than
  15.5 mag (linear scale, contours trace density in steps of 20, 30
  ... 90 percent of maximum density). The density is given in units of
sources per arcsec$^2$.}
\label{Fig2DDensityA}
\end{figure}
We fitted the projected early type density with a power law
$\rho_{proj.} \propto R^{\beta_{1''}}$, excluding the inner 1''. But
while providing a value that agrees well with previously published
results \citep{genzel2003,paumard2006,lu2008}, this single power law
does not provide a very good fit to our data. For comparison, we also
fitted the projected early type density with a broken power law with a
break at 10''. This minimizes the deviations from the data, but it
introduces a break at 10'' that we cannot explain.    
\begin{eqnarray}
\beta_{1''} = -1.49 \pm 0.12\nonumber\\
\beta_{1-10''} = -1.08 \pm 0.12\nonumber\\
\beta_{10-20''} = -3.46 \pm 0.58\nonumber
\end{eqnarray}
It has to be kept in mind that the absolute number of
early type stars in the outer regions is much lower than further
towards the center, so the uncertainties are considerably larger. The
value for R $>$ 1'' agrees reasonably well with the values determined
by \cite{genzel2003,paumard2006,lu2008}. At R $<$ 1'', the early
  type density is lower than expected from extending the power law
  inwards. This is the sign of the inner edge of the clockwise disk
  reported by \cite{paumard2006} and Lu et al. (2008). In this densest
  part of the cluster, the stellar surface density is probably also
  underestimated because source confusion will lead to incompleteness
  of the data (we estimate that completeness of sources at $mag_{K}
  \leq15.5$ drops to $\sim80\%$ within $0.5''$ of Sgr\,A* ). Outside
of 1'', we detect a number of early type stars (312 over the whole
field compared to 90) that is by a factor of $\sim$3 higher than
in previous works. Their density profile is similar to that of the
previously reported early-type sources. The power law in
  the inner few arcseconds becomes flatter with the additional
  early-type stars included, but agrees within the uncertainties with
  the previously reported values. Towards the edges of the cluster
($\sim$10-20''), we observe a steeper density profile than
\cite{paumard2006}. It may be possible that a different density law
applies outside of 10'', but the cause for such a phenomenon is
unknown. It has to be considered, however, that the statistics for
such a small number of stars at larger distances are not very reliable
any more. In addition, two different cutoffs were used for sources
inside and outside of 12''. This can also lead to a bias here in the
way that the early type density is underestimated outside of 12'' and
overestimated within that distance to Sgr A*. But this effect should not
influence the density in the order of magnitude observed here which
leads to the different slopes of the fitted power
laws.\\
\begin{figure}[!b]
\centering
\includegraphics[width=\textwidth,angle=-90, scale=0.4]{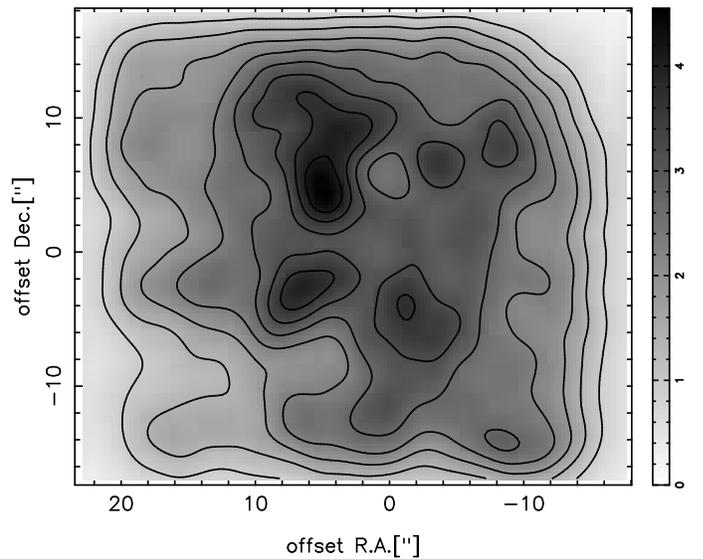}
\caption{\small Stellar surface density of late type stars brighter than 15.5 mag (linear scale, contours trace density
  in steps of 20, 30 ... 90 percent of maximum
  density). The density is given in units of
sources per arcsec$^2$.}
\label{Fig2DDensityL}
\end{figure}
The relative stellar density of the sources rated as too noisy (see
Fig.\ref{FigPDensityN} appears to be flat over an inner region with a
radius of $\sim$15''. Outside of 15'', the ratio increases, but that
can be expected since the quality of the photometry decreases
  somewhat toward the edge of the FOV. This is probably due to the
  rectangular dither pattern used that leads to shallower
  integration toward the edges of the FOV.\\ This gives
further evidence that our criteria for excluding noisy sources as well
as the local calibration we apply are justified, since an even
distribution of noisy sources can be expected from a well calibrated
dataset. This is the case here, and it means that
the exclusion of noisy sources does not lead to a bias in our surface
density profiles.\\
Fig.\ref{Fig2DDensityA} shows the two-dimensional density distribution
of all stars brighter than mag$_K = 15.5$ and Fig.\ref{Fig2DDensityL}
that of the late type stars of the same magnitude range, while
Fig.\ref{Fig2DDensityE} shows the density of the early type stars in
the same way. These maps have been smoothed with a $\sim$4''
Gaussian. While the late type stars show a similar distribution as the
stars in the cluster viewed as a whole (with the exception of the
central few arcseconds), the early type stars are concentrated in the
center. This result is not surprising since the same can be seen from
the azimuthally averaged density. The area in the immediate vicinity of IRS 7
shows a significantly lower stellar density in all our maps, because the presence of this
extremely bright source impedes the detection of other stars close to
it.\\
\begin{figure}[!t]
\centering
\includegraphics[width=\textwidth,angle=-90, scale=0.4]{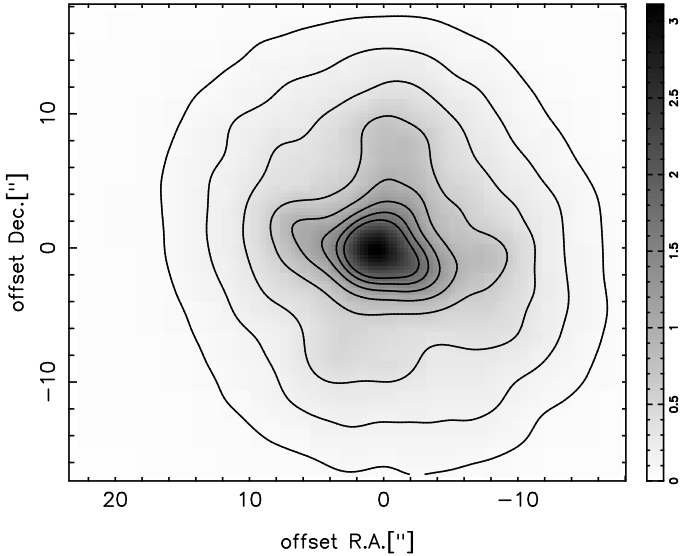}
\caption{\small Stellar surface density of early type stars brighter than 15.5 mag (logarithmic scale, contours trace density
  in steps of 2.5, 5, 10, 20 ... 90 percent of maximum
  density). The density is given in units of
sources per arcsec$^2$.}
\label{Fig2DDensityE}
\end{figure}
The peak of the early type
density is, as expected, located close to the position of Sgr A*. The
distribution of the early type stars appears close to circularly
  symmetric, but indicates extensions along N-S and E-W. An apparent
  concentration of early -type stars along these directions can also
  be seen in Fig.\,1 of \cite{bartko2008}.\\
 The density distribution of the late type stars
  clearly shows the relative lack of late type stars in the very
  center. The map also shows a
  correlation with the extinction map (see extinction map provided in
  \cite{schoedel2007} and Fig.\ref{FigExtmap}). Areas of higher
  extinction show a lower density of stars. This effect is not visible
  in the early type density. But in the areas with higher extinction
  (e.g. 5'' NW of Sgr A*), the early type density is too low for this effect to be
  relevant, although a trace of it can be seen in the slight dip in
  the early type density in Fig.\ref{FigPDensity}. The extinction has two separate effects on the projected
  density: one effect is that sources behind a lot of extinction
  appear fainter, so they might be excluded by a simple magnitude
  cutoff. This effect has been corrected here by using extinction
  corrected magnitudes. The other problem is that extinction also
  impedes the detection of sources, especially in filters with lower
  image quality. This effect of the extinction on the stellar density
  has not been calculated here and was not compensated.
\subsection{Evidence for giant depletion in the center}
\label{SectGiants}
\cite{figer2003} examined the radial
velocities of 85 cool stars in the GC (mostly M and K giants) and
found dynamical evidence for a flattened distribution of late type
stars within 0.4 pc ($\sim$ 10''). As \cite{figer2003} and
\cite{zhu2008} point out, the flat projected surface density profile
of the late-type stars implies in fact a {\it hole/dip} in the
3-dimensional distribution of the late-type stars.\\
\cite{schoedel2007} described the total population of the central
parsec with a broken power law (break radius R$_{break} = 6''.0 \pm
1''.0$, $\beta = -0.19  \pm 0.05$ within R$_{break}$ and $\beta = -0.75
\pm 0.10$ outside of the break radius) for a magnitude limit of 17.75
and using completeness corrected data. This does not allow a
comparison of absolute densities to our findings, since
that dataset goes much deeper than ours, but the trend can be
confirmed here. It also has to be considered that these results were
obtained on the entire population and not separated into early and
late type stars.\\
\begin{table}[!t]
\caption{\small Power law indices for late type stars and all
  classified stars, separate fit to sources inside and outside of 6.0''}
\label{TabBPL}
\centering       
\begin{tabular}{l r r}
\hline\hline
 & $\beta_{inner}$ & $\beta_{outer}$\\
\hline
all stars & -0.22 $\pm$ 0.11 & -0.86 $\pm$ 0.08\\
late type stars & 0.17 $\pm$ 0.09 & -0.70 $\pm$ 0.09\\
\hline
\end{tabular}
\end{table}
We fitted broken power laws to the projected densities of the late
type stars and all classified stars (see Tab.\ref{TabBPL}).
We find the same break radius of 6.0$\pm$1.0'' as \cite{schoedel2007},
and the power law indices for the total population also agree with the
values given in that work (-0.22$\pm$0.11 for the inner region,
-0.86$\pm$0.06 for R$>$6'' in our data). What is new here, however, is the
possibility to obtain separate values for the late type population
alone. These values give an even stronger support to the proposed
hole/dip in the center: we find a power law index of -0.70$\pm$0.09
for the outer region, while the inner region even shows a decline
towards the center (0.17$\pm$0.09).\\ 
 \begin{table}[!b]
\caption{\small KLF power law indices of the different classified stellar types, calculated for the entire cluster resp. the inner 7''}
\label{TabKLFslopes}
\centering       
\begin{tabular}{l c c}
\hline\hline
type & complete cluster & r $<$ 7''\\ 
\hline
all & 0.26 $\pm$ 0.01 & 0.21 $\pm$ 0.02\\ 
late & 0.31 $\pm$ 0.01 & 0.27 $\pm$ 0.03\\
early & 0.14 $\pm$ 0.02 & 0.13 $\pm$ 0.02\\             
\hline
\end{tabular}
\end{table}
This is a very interesting result, since it shows that the previously
observed flattening of the density profile of the total population
\citep{genzel1996,figer2003,genzel2003,schoedel2007,zhu2008} is an even stronger
feature in the late type population. We can therefore assume that the
stellar population in the innermost $\sim$0.2 pc is indeed depleted
not only of bright giants, but also of fainter giants down to our
magnitude limit of 15.5 mag. Several causes for this have been proposed: \cite{dale2008}
offered an explanation for the under-density of late type stars within
1'', claiming that collisions with stellar mass black holes and main
sequence stars prevent 1-2 M$_{\odot}$ giants to evolve so that they
are not visible in the K band. Their simulations cannot explain the
lack of brighter and fainter giants.\\
\cite{freitag2008} derived collision probabilities for bright giants in the
GC (see esp. their Fig.1). They determined that nearly all massive stars
within 0.1 pc almost certainly suffer from collisions during
their time on the giant branch.\\
\begin{figure*}[!t]
\centering
\includegraphics[width=\textwidth,angle=-90, scale=0.7]{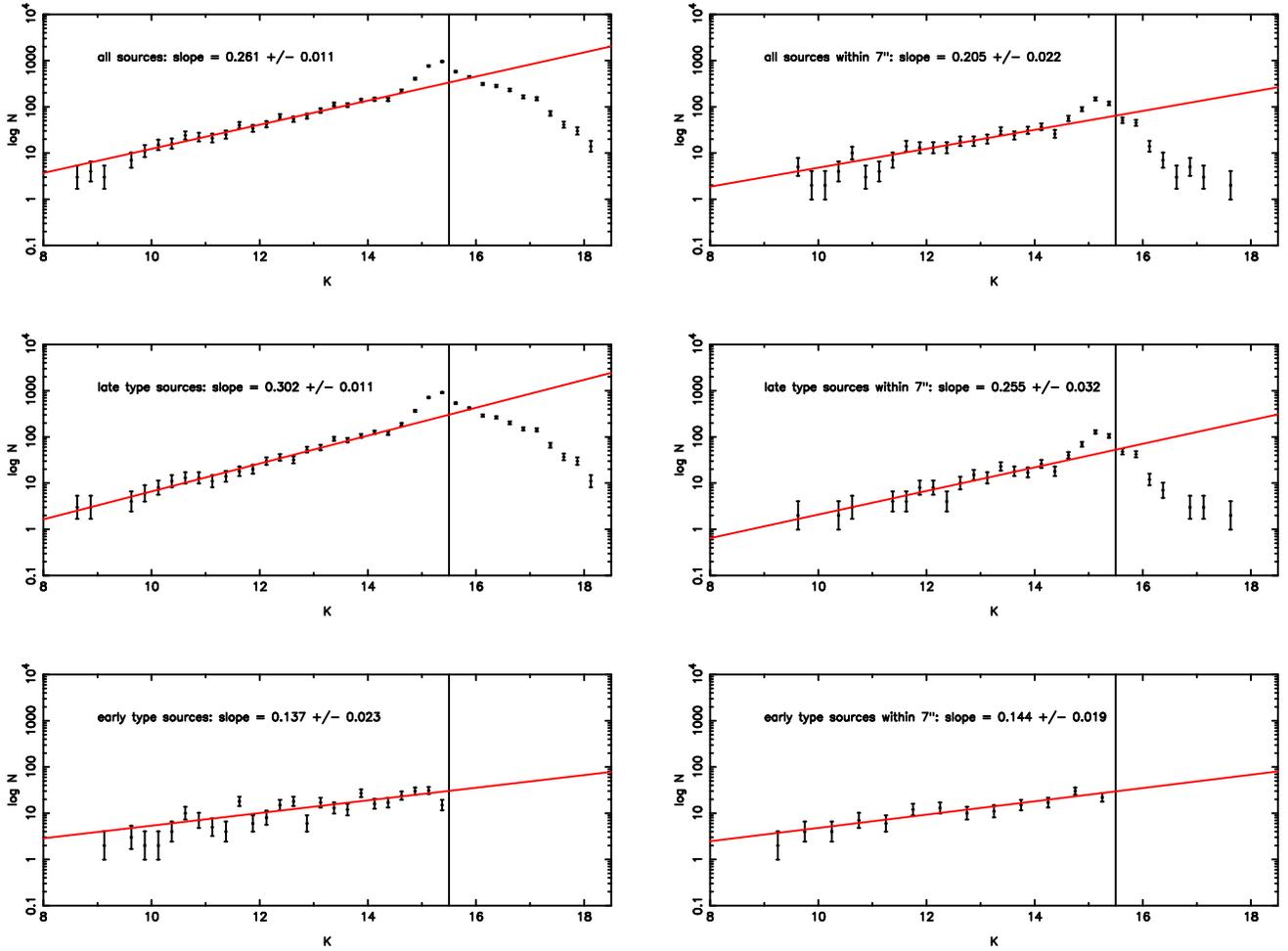}
\caption[K band luminosity function of early and late type
  stars]{\small Luminosity function of stars in the central
  parsec. Each data-point represents the center of a magnitude bin
  (0.25 mag wide bins, resp. 0.5 mag in lower right frame). Upper left: all stars, the HB/RC bump at $\sim15.0$ mag is
  obvious. Central left: only late type
  stars, the HB/RC bump is visible. Lower left: only early type stars,
  much flatter luminosity function, no HB/RC bump. Upper right: all
  stars within 7''. Central right: late type stars within 7''. Lower right:
  early type stars within 7''. The luminosity functions in the central
  few arcseconds of the cluster appear to have a flatter slope than
  that of the complete cluster, with the exception of the early type
  stars: here the slopes are almost the same.}
\label{FigLumfct}
\end{figure*}
\cite{merritt_szell2006} offered yet another explanation: the infall
of a second SMBH would destroy the stellar density cusp present around
Sgr A*, which would then be built up again in a time-frame of several
Gyrs. This process can lead to a practically flat density profile,
similar to the one observed here. 
\subsection{K band luminosity function}
\label{sect_lumfct}
The general K band luminosity function (Fig.\ref{FigLumfct}) agrees well with the one
presented by e.g \cite{genzel2003,schoedel2007}. The red clump can
clearly be made out at the expected magnitude of
$\sim$15.0-15.25. The
15.5 mag limit of the observations does not have a significant effect
on this feature, since this limit only applies to the separation
of early and late type stars, while the photometric completeness limit
lies at $\sim$ 16 mag.\\
The luminosity functions can be described to the first order by a
power law (fitted to the area between 9.0 and 15.5 mag for the
early type stars, resp. 14.5 to exclude the red clump in the other
plots):
\begin{equation}
\frac{d log N}{dK} = \beta
\end{equation}
The power law indices that resulted from this fit are shown in Tab.\ref{TabKLFslopes}.
\cite{alexander_sternberg1999,tiede_frogel_terndrup1995,zoccali2003}
measured the power law slope of the bulge population of the milky way
several degrees from the center as $\beta \sim 0.3$, while
\cite{figer2004} give the same value for the KLF on 30 pc scales
around the GC, claiming that this value agrees very well with the
theoretical KLF of an old stellar population, reflecting the rate of
evolution of stars along the red giant and asymptotic giant
branch. \cite{genzel2003} give a value of $\beta = 0.21 \pm 0.02$ for
the central 9''. It is expected that for greater distances to the
center, the slope value approaches that of the bulge.\\ The power law
fitted here for the total population of the central parsec is flatter
than the one attributed to the bulge population, while our value
fitted for the inner 9'' matches the value of 0.21 given by
\cite{genzel2003}.\\ The individual KLFs for the late and early types
give the reason for this deviation from the bulge power law: since the
central parsec (and even more the central few arcseconds) contain a
significant number of early type stars and their KLF has a much
flatter slope ($\beta_{early} = 0.14 \pm 0.02$), the resulting power
law is also flatter than that of the late type population alone. The
power law fitted to the KLF of only the late type stars ($\beta_{late}
= 0.31 \pm 0.01$) agrees very well with the one observed in the
bulge.\\
\begin{figure*}[!t]
\centering
\includegraphics[width=\textwidth,angle=-90, scale=0.7]{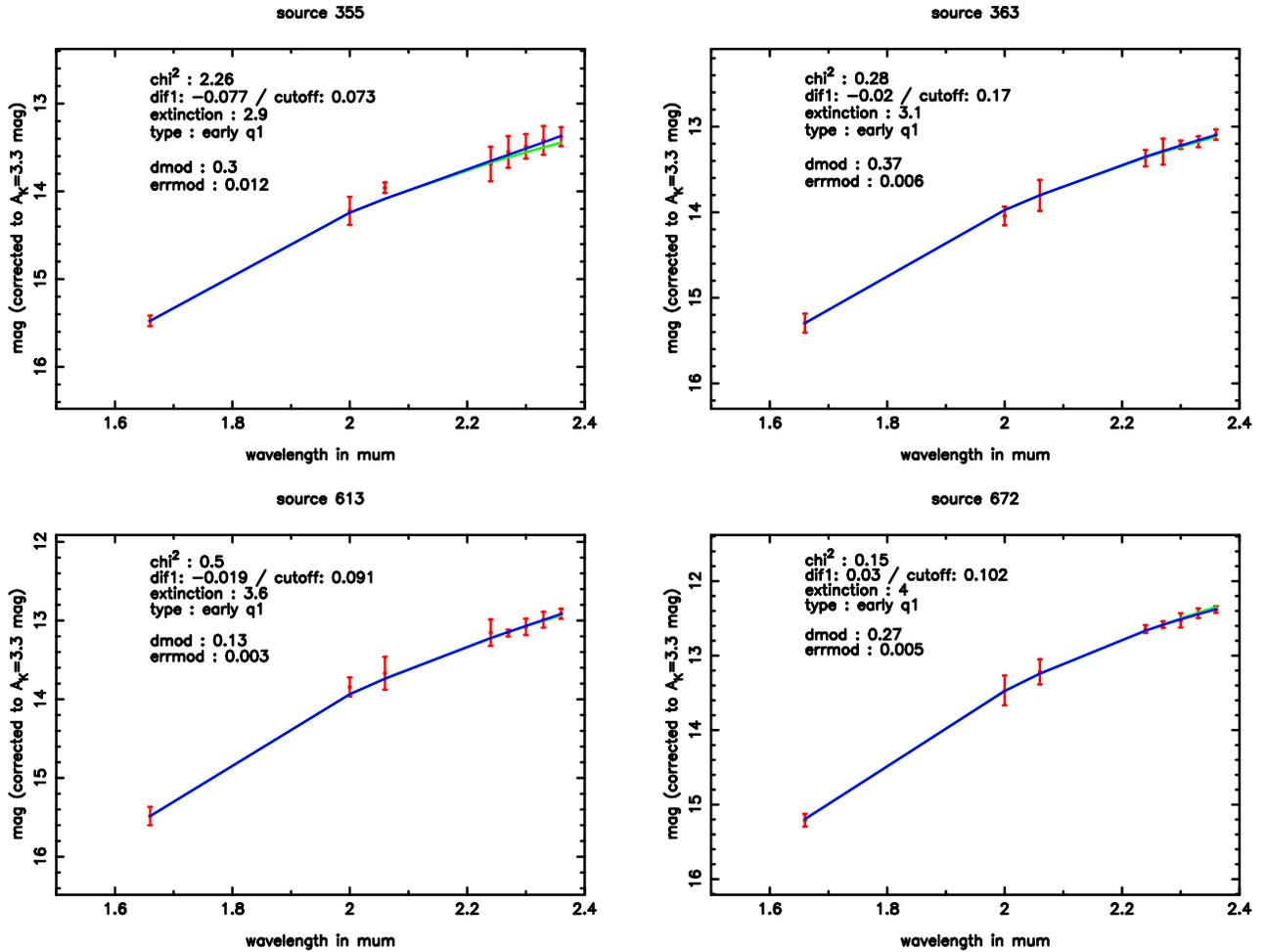}
\caption{\small SEDs of stars fitted as early type outside of
  0.5 pc. We consider these four examples to be very likely candidates
  for actual early type stars. The source in the upper right frame
  does not show a CO absorption feature, but still it has been
  labeled as a late type source in
  \cite{maness2007}. \cite{paumard2006} seem to list it as an early
  type however. The other sources listed in Tab.\ref{TabEtypesFarout}
  with quality A have similarly smooth SEDs, while quality B and C sources
  are considerably more noisy.
  }
\label{FigEtypesFarout}
\end{figure*}
Agreeing with
\cite{lebofsky_rieke1987,blum1996,davidge1997,genzel2003}, our data
point to the population in the central parsec being an old stellar
population with an admixture of a young, bright component.\\ The
fitted slope value for the inner 9'' of 0.21 differs from the one for
the total cluster, but otherwise, the shape of the KLF there is very
similar. This difference is due to the flatter slope of the late type
KLF in this region, while that of the early type KLF stays the
same.\\ \cite{paumard2006} also presented a KLF for the early type
stars in the disks. They also find a flat KLF, similar to the one
presented here. The significant difference is the greater
  sensitivity we achieved here.
\cite{paumard2006} note that their
KLF has a spectroscopic completeness limit of $\sim$13.5-14 mag,
compared to 15.5 mag here. The fact that the flat slope of the early
type KLF can be observed down to that magnitude is an important new
result and it strengthens assumptions of a top-heavy mass function
since it improves the statistic relevance of the observed flatness.
\subsection{Extinction}
\begin{figure*}[!t]
\centering
\includegraphics[width=\textwidth,angle=-90, scale=0.35]{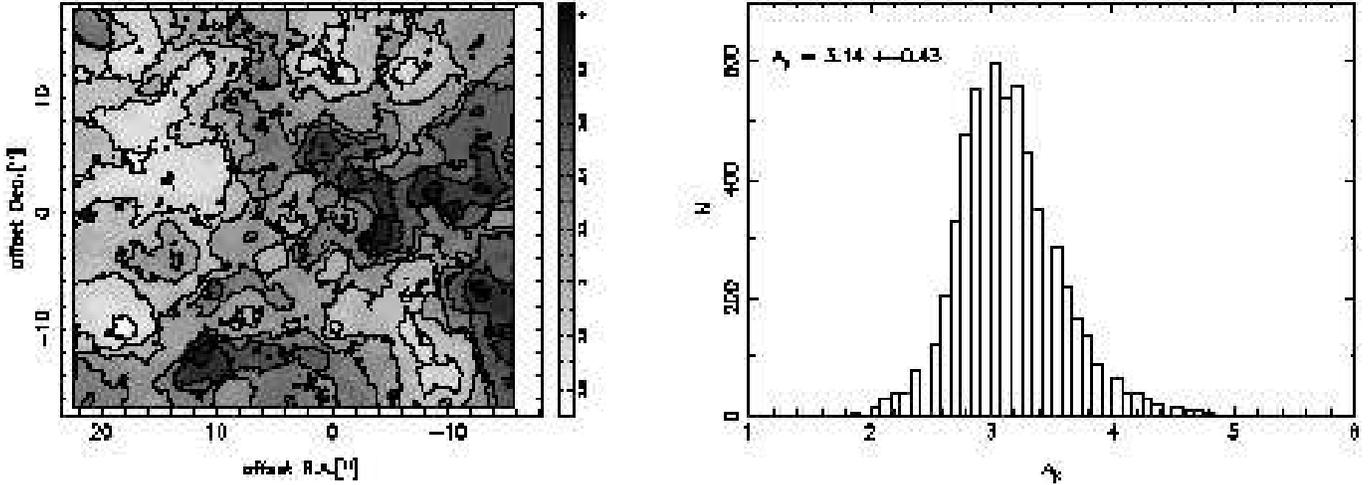}
\caption{\small Left: Extinction map of the central parsec computed from
  individually fitted extinction values. Features like the mini-cavity
  and the mini-spiral are visible. Right: Distribution of extinction
  values in the central parsec of the GC. The mean value and the
  variation agree with previously published results. Note that the
  Draine(1989) extinction law was used here and that a different law
  might apply to the GC. This may lead to a systematic offset as large
  as A$_K$ = 0.5 mag, which should be taken into account as a possible
  systematic uncertainty when interpreting these values.}
\label{FigExtmap}
\end{figure*}

\begin{figure*}[!t]
\centering
\includegraphics[width=\textwidth,angle=-90, scale=0.7]{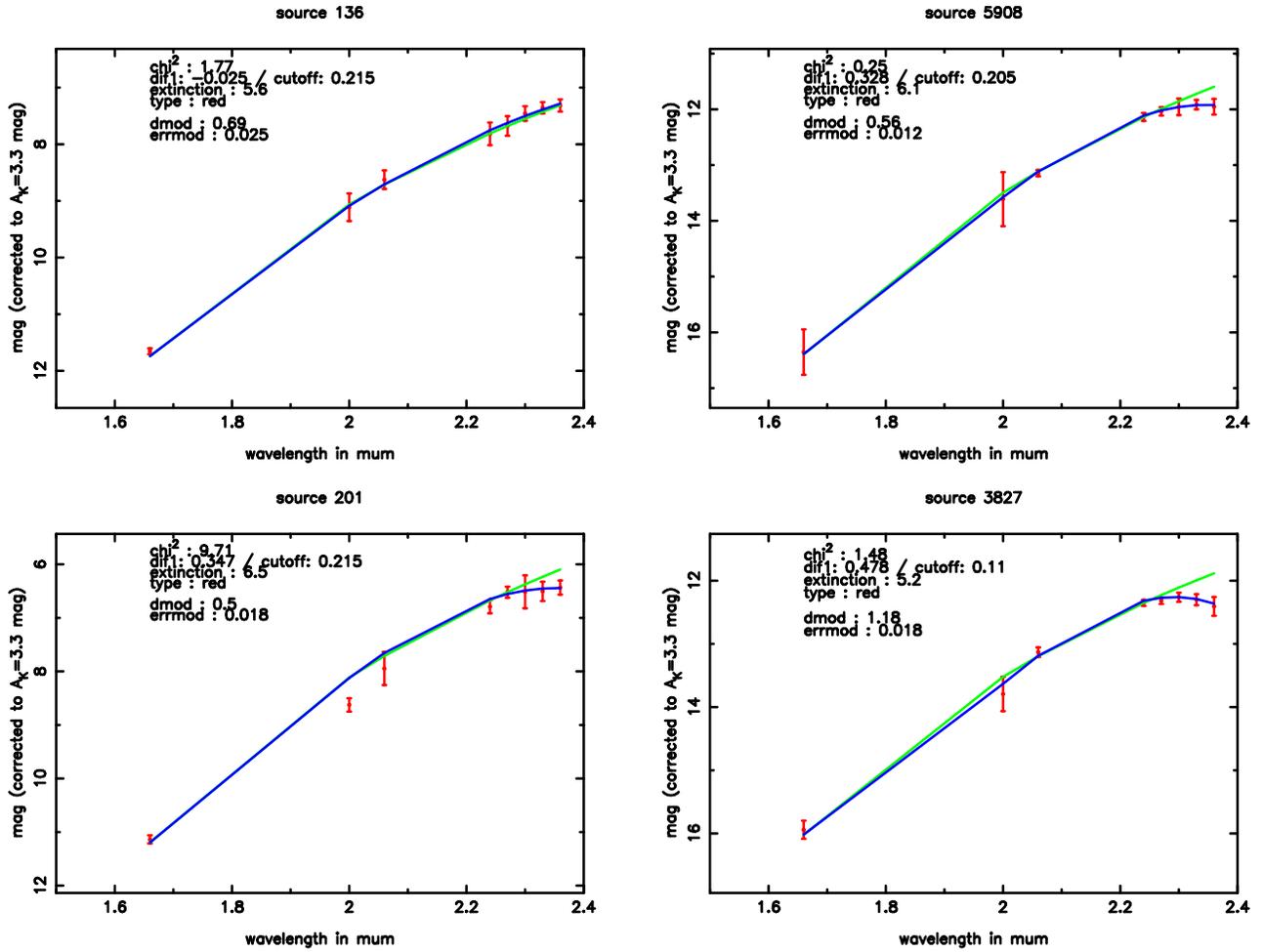}
\caption{\small SEDs of
    Extremely Red Objects. Upper left: source NW of Sgr
    A*, no CO feature, but strongly reddened. Upper right: source NW
    of Sgr A*, here a CO feature is visible. This source is located
    near a local maximum of extinction, so it may just be a normal
    late type star. Lower left: source located in the northern arm of
    the mini-spiral. Lower right: source located in the far SE of Sgr
    A*, also in a region of very high extinction. The CO feature
    suggest a late type star, maybe towards the back of the cluster.}
\label{FigEROs}
\end{figure*}

The average extinction of 3.3 mag towards the GC is well known
(e.g. \cite{scoville2003}), but this value varies on
small scales by up to 1.5-2 mag \citep{schoedel2007} and poses a significant problem for
the reliable measurement of apparent magnitudes, colors and intrinsic
reddening of stars. \cite{schoedel2008} are presenting an
extinction map based on H-K colors obtained from the same data that we
are using here, but the individual extinction values obtained here can
also be used to produce an extinction map. Fig.\ref{FigExtmap} shows
this map. It agrees well with \cite{schoedel2007}, except that we find
higher overall extinction values.  The extinction map produced in
  this work is expected to be more reliable since we use many
wavelength bands, apply a local calibration and distinguish between
hot and cool sources. A histogram of the measured extinction
  values is shown in Fig.\ref{FigExtmap}: The distribution is similar to
  a Gaussian with a mean value and standard deviation of 3.1 $\pm$
  0.4. This agrees well with previously published results
  \citep{scoville2003} and also with
  \cite{schoedel2008}. The asymmetry of the histogram is due to the
  exclusion of foreground stars.
\subsection{Early type stars outside of 0.5 pc}
\label{sect_etypes}
In addition to not only identifying the well known early type stars in
the center of the cluster, but indeed more than tripling the number of
early type candidates in the central 0.5 pc, we can also report the
identification of 35 early type candidates more than 0.5 pc
($\sim$12.9'') from Sgr A*. If we apply our previously derived
uncertainty of $\sim$9\% (see \S \ref{SectCompref}), this leads us to
a number of $35 \pm 3$ early type candidates outside of the inner
region. This number is most likely underestimated, due to the stricter
cutoff criterion used in this region (see \S \ref{SectCutoff}). Within the observed area, the distribution of
these sources appears to be fairly isotropic, although one has to be
cautious here due to the the asymmetry of the observed region and the
small number of candidates. The nature of these candidates needs to be
confirmed with spectroscopic observations, since our method can only
provide a first estimate for the type.\\
We obtain an average
  density of early type stars of $(4.6 \pm 0.4) \times 10^-2$
  sources/arcsec$^2$ at $R > 12.9''$.\\
The detection of early type
stars this far out in significant numbers is a new result: \cite{paumard2006} reported no
early type stars outside of the central 0.5 pc, citing an 1 $\sigma$
upper limit of $\sim$10$^{-2}$ OB stars per arcsec$^2$ outside of 13''
deduced from SINFONI data, but referring to an unpublished source. The
difficulty in identifying these stars in existing data is pointed out
by \cite{trippe2008}, who mention very limited coverage of the central
cluster and data gathered with several instruments and in different
epochs that is very difficult to compare due to different pixel
scales, Strehl ratios and completeness. These obstacles have probably
impeded the detection of early type stars in the outer region of the
central cluster until now. Our new method can provide targets for
spectroscopic confirmation observations, allowing for a broader search
for early type candidates over a large area without the need to cover
the whole area with integrated field spectroscopy. Finally, we
  would like to point out that we do know that at least one early type
  star exists outside of the central 0.5\,pc, as has been confirmed by the
  results of \cite{geballe2006} on IRS\,8.\\ 
As Fig.\ref{FigPDensity} shows, there are two possible fitting
solutions for the early type density. While the single power law
fitted to the projected density of early type stars agrees with the
previously published power law, the slopes of the broken power law
show a significant difference in the outer and inner region. If a
single power law distribution can be assumed, this would indicate that
these stars merely represent the continuation of the disk and the off-disk
population. But if the much steeper decline outside of 10'' is indeed
a significant and real feature, this might point to a change in the
population respectively so far unknown effects on the density
distribution.\\
\begin{table}[!b]
\caption{\small Early type stars
  detected outside of 0.5 pc. {\em x} and {\em y} denote the position of the star
  in arcsec relative to Sgr A*, $mag_{K,ext}$ the extinction
  corrected K band
  magnitude and $A_K$ the extinction in the K band. {\em Quality}
  indicates the confidence in the identification after an additional
  visual inspection, with A as the highest confidence and C as the lowest.}
\label{TabEtypesFarout}
\centering       
\begin{tabular}{r r r r l r}
\hline\hline
x('') & y('')& mag$_{K,ext}$& A$_K$ & CBD & quality\\
\hline
-8.15 & 13.50 & 10.7 & 3.3 & 0.07$\pm$0.04 & C\\ 
-9.59 & -15.33 & 12.5 & 3.1 & -0.05$\pm$0.01 & B\\
15.74 & 12.87 & 13.1 & 2.8 & 0.03$\pm$0.03 & B\\
-8.37 & -11.27 & 13.9 & 2.7 & -0.095$\pm$0.019 & C\\
-8.64 & -10.74 & 13.5 & 3.0 & -0.003$\pm$0.003 & A\\
-10.57 & -15.90 & 13.0 & 3.3 & 0.05$\pm$0.01 & B\\
18.92 & 0.28 & 13.3 & 3.2 & 0.02$\pm$0.01 & B\\
-10.62 & -9.76 & 13.2 & 3.3 & 0.008$\pm$0.017 & A\\
-10.81 & -10.51 & 14.1 & 3.0 & 0.01$\pm$0.01 & B\\
13.05 & 14.29 & 13.3 & 3.5 & 0.04$\pm$0.02 & B\\
7.39 & 12.29 & 13.3 & 3.5 & 0.03$\pm$0.01 & A\\
13.13 & 5.08 & 14.6 & 2.8 & -0.07$\pm$0.04 & B\\
19.58 & -16.54 & 12.7 & 4.0 & 0.032$\pm$0.005 & A\\
4.43 & 16.11 & 13.9 & 3.4 & 0.028$\pm$0.007 & B\\
9.43 & 9.29 & 13.6 & 3.5 & -0.008$\pm$0.018 & B\\
18.70 & 4.44 & 14.0 & 3.3 & -0.08$\pm$0.03 & B\\
-10.86 & 7.20 & 13.3 & 3.8 & 0.05$\pm$0.02 & B\\
11.30 & 8.47 & 14.0 & 3.4 & 0.018$\pm$0.004 & A\\
17.69 & 3.01 & 14.0 & 3.4 & 0.03$\pm$0.01 & C\\
10.91 & 9.44 & 11.9 & 4.7 & -0.07$\pm$0.03 & B\\
-9.05 & -11.77 & 15.2 & 2.9 & -0.089$\pm$0.005 & A\\
-12.75 & 9.77& 12.8 & 4.4 & 0.05$\pm$0.02 & B\\
17.48 & 11.36 & 13.7 & 3.9 & 0.036$\pm$0.008 & A\\
-4.20 & 14.06 & 14.9 & 3.2 & -0.08$\pm$0.02 & B\\
17.51 & -8.95 & 14.8 & 3.4 & -0.07$\pm$0.01 & A\\
9.89 & -14.14 & 13.2 & 4.4 & 0.04$\pm$0.03 & B\\
14.25 & -5.64 & 14.5 & 3.8 & -0.033$\pm$0.005 & A\\
12.55 & -4.19 & 13.2 & 4.7 & 0.03$\pm$0.02 & A\\
-15.25 & 6.68 & 14.1 & 4.1 & -0.09$\pm$0.06 & B\\
-13.13 & -12.44 & 15.1 & 3.6 & -0.079$\pm$0.004 & A\\
-5.14 & -15.75 & 15.4 & 3.4 & -0.13$\pm$0.02 & C\\
4.49 & -15.53 & 15.2 & 3.6 & -0.14$\pm$0.03 & B\\
-1.16 & 16.91 & 13.8 & 4.5 & -0.06$\pm$0.02 & B\\
0.31 & -15.30 & 15.1 & 3.7 & -0.11$\pm$0.03 & B\\
8.86 & 9.55 & 15.4 & 3.5 & -0.12$\pm$0.02 & C\\
\hline
\end{tabular}
\end{table}
\section{Conclusions}
\label{SectDiscussion}
Our newly developed method has confirmed several previously obtained
results and has proven to be able to classify sources of known type
reliably. It cannot compete with spectroscopic identifications of
individual sources, but due to the ability to classify a large number
of sources with a relatively small need for observation time, it can
provide important statistical information about a stellar population
which can later be refined by observing the early type candidates with
spectroscopic methods. Our analysis extends the sensitivity limit
  of stellar classification by about 1.5\,magnitudes compared to
  previous work, providing thus a statistically stronger basis
for conclusions on the stellar population.\\ The following results
could be obtained:
\begin{enumerate}
\item A larger number of early type candidates has been detected than
  in any previous study, 312 sources compared to the known 90. We were
  able to identify known early type sources with 87\% accuracy, while
  96\% of the known late type sources were classified correctly. The
  different percentages stem from the different selection criteria of the
  reference sources in the cited publications: \cite{maness2007} already selected their late
  type sources by the presence of sufficiently deep CO band heads, which
  is the same feature that our method makes use of. The early type
  sources published by \cite{paumard2006} were selected based on
  narrow emission lines that are not visible at our spectral
  resolution.\\
  These high rates of correct classifications gives us high confidence
  in our results, but the new detections need to be confirmed
  spectroscopically, especially the early type candidates.
\item The spatial distribution of the early type stars follows a
  power law, with $\beta_{1''} = 1.49 \pm 0.12$, which is flatter than
  the values of the value of $\beta = -1.8$ we computed from the
  data published by \cite{paumard2006,lu2008}, but still compatible at the $3\sigma$
  limit. It is also possible to fit the early type density with a
  broken power law with $\beta_{1-10''} = -1.08 \pm 0.12$ and
  $\beta_{10-20''} = -3.46 \pm 0.58$. This broken power law fits the
  density distribution better than the single power law, but it
  remains unknown what causes the density drop we observe at
  $\sim$10''. The significance of this feature should be examined by
  further observations that extend to larger distances from Sgr A*.
\item We confirm the previously reported flat projected surface
  density profile of the late type stars in the innermost arcseconds
  with much higher source numbers and therefore significantly improved
  statistics. The flat (or even inversed) surface density of late type
  stars must imply a central {\it dip or hole} in their 3D
  distribution. This together with the steep early type density
  profile explains the observed drop in CO absorption. We would
    like to emphasize that this result signifies a change in our
    perception of the nuclear star cluster. Separating the early- and
    late-type population shows clearly that the GC cluster does {\it
      not} have a {\it cusp} (see discussions in \cite{genzel2003,schoedel2007}). To the contrary, the late-type, old
    stellar population, which appears to make up the vast majority of stars in the GC cluster, shows a flat or even slightly inverted
    power-law in projection within about $0.2$\,pc from Sgr\,A*. This
    means that there is some kind of {\it hole} in the late-type
    population near the center, as has been pointed out by \cite{figer2003}. The exact cause is still not understood, but various
    explanations have been suggested. One often discussed possibility
    is the destruction of the envelopes of giant stars by collisions
    with main-sequence stars, post-main sequence stars, and stellar
    remnants in the dense environment near Sgr\,A*
  \citep{rasio_shapiro1990,davies_benz1991,genzel1996,alexander1999,davies1998,bailey_davies1999,dale2008}.
  Another possible explanation for the non-existent cusp of
    late-type stars is that the cusp may have been destroyed by the
    infall of a second black hole \citep{merritt_szell2006} and that
    there may not have been sufficient time yet to re-grow the cusp.
\item The late type KLF has a power law slope of 0.30$\pm$0.01. This
  resembles closely the KLF that has been measured for the bulge of
  the Milky Way. This is surprising considering that the
    nuclear star cluster is probably a dynamically separate entity
    from the bulge (see \cite{boker2008} for an overview of the properties of nuclear
    star clusters in galaxies). It implies a similar star formation
  history for the NSC and the bulge. The early type KLF has a much flatter
  slope of ($0.14 \pm 0.02$). The fact that the flatter early type KLF
  could be confirmed down to our magnitude limit of 15.5 mag is an
  important extension of previous works that had a completeness limit
  of $\sim$13-14 mag. \cite{paumard2006} claimed that their flat early
  type KLF agreed best with stellar evolution models using a top-heavy
  initial mass function (IMF), but did not give a value for the slope
  of the KLF itself (Fig.13 in that work). Our early type KLF
  seems to show a very similar shape down to our magnitude limit and
  thus also appears to support the proposed top-heavy IMF, but an accurate
  comparison is not possible due to the lack of a value for the slope
  in \cite{paumard2006}.  
\item Early type stars have been detected outside of 0.5 pc, in a
  density that still agrees with the power law density profile in the
  inner region, so these sources could well be part of the known
  disk/off-disk population, if the single power law is valid. If the
  edge we observe at $\sim$10'' is a real feature, this may point to a
  more complicated situation. This result is of course pending
  spectroscopic confirmation. We find a larger number of early type
  sources in the whole cluster with the density distribution following
  the same power law as in \cite{paumard2006}. This means that our
  power law still agrees with our early type density outside of 0.5 pc
  that is higher than the upper limit provided in that work.
\item Both foreground stars and strongly reddened objects could be
  detected or excluded easily due to their fitted
  extinction. Unfortunately not all known bow-shock and mini-spiral
  sources could be observed, due to position and photometric uncertainties.   
\end{enumerate}
It is still not decided which one of the two main scenarios
  serves best to explain the presence of the early type stars in the
  central half parsec, in situ formation or infall and dissolution of
  a cluster formed at several parsecs distance from the GC. Recent
publications lean towards favoring the scenario of {\em in-situ} star
formation (e.g. \cite{paumard2006,nayakshin_sunyaev2005,bartko2008}. Our results also seem to agree best with
this model: the steep power law decline of the projected early type
density that agrees much better with the R$^{-2}$ power law expected
for that scenario than with the R$^{-0.75}$ of the {\em in-spiraling
  cluster}. It is important to note that the cluster infall
  scenario predicts that less massive stars are stripped from the
  cluster at larger distances \citep{guerkan_rasio2005}. Our
  analysis probes, for the first time, the density of late-O/early
  B-type stars in the entire central parsec. These stars are less
  massive than the early-type stars reported from spectroscopic
  observations \citep{paumard2006,bartko2008}. Although we
  find a somewhat flatter power-law for the density of the early-type
  stars, the improved statistics still imply a steep decrease of
  the early-type stellar density with distance from Sgr\,A*. This
  supports the {\it in situ} scenario, but it is not possible to
  clearly rule out or confirm one scenario based on the current
  data.\\
Further observations
that cover a larger area should be undertaken in the future, as well
as spectroscopic confirmations of our new early type
candidates. \cite{lu2008} suggest covering at least the inner 5 pc of
the GC, and while observations of such an area require an extreme
amount of time with an instrument like SINFONI, our method may be the
key to explore large areas like this for features like tidal tails of
an in-falling cluster or a continuation of the early type population of
the central parsec.
\begin{acknowledgements}
We are grateful to all members of the NAOS/CONICA and the ESO PARANAL
team. R. Sch\"odel acknowledges support by the Ram\'on y Cajal
programme by the Ministerio de Ciencia e Innovaci\'on of the
government of Spain. We would also like to thank the referee for his
helpful comments.
\end{acknowledgements}

\end{document}